\documentclass[aps,onecolumn,showpacs,letterpaper]{revtex4}

\usepackage{graphicx}
\usepackage{dcolumn}
\usepackage{amsmath}
\usepackage{longtable}

\begin{document}

\title{Nuclear deformation and neutrinoless double-$\beta $ decay of $^{94,96}$Zr, $%
^{98,100}$Mo, $^{104}$Ru, $^{110}$Pd, $^{128,130}$Te and $^{150}$Nd nuclei
in mass mechanism\\
}
\author{K. Chaturvedi$^{1,2}$, R. Chandra$^{1,3}$, P. K. Rath$^{1}$, P. K. Raina$^{3}
$ and J. G. Hirsch$^{4}$}
\affiliation{$^{1}$Department of Physics, University of Lucknow,
Lucknow-226007, India\\
$^{2}$Department of Physics, Bundelkhand University, Jhansi-284128, India\\
$^{3}$Department of Physics and Meteorology, IIT Kharagpur-721302, India\\
$^{4}$Instituto de Ciencias Nucleares, Universidad Nacional Aut\'{o}noma de
M\'{e}xico, A.P. 70-543, M\'{e}xico 04510 D.F., M\'{e}xico} 

\date{\today}

\begin{abstract}
The $\left( \beta ^{-}\beta ^{-}\right) _{0\nu }$ decay of $^{94,96}$Zr, $%
^{98,100}$Mo, $^{104}$Ru, $^{110}$Pd, $^{128,130}$Te and $^{150}$Nd isotopes
for the $0^{+}\rightarrow 0^{+}$ transition is studied in the Projected
Hartree-Fock-Bogoliubov framework. In our earlier work, the reliability of
HFB intrinsic wave functions participating in the $\beta ^{-}\beta ^{-}$
decay of the above mentioned nuclei has been established by obtaining an
overall agreement between the theoretically calculated spectroscopic
properties, namely yrast spectra, reduced $B(E2$:$0^{+}\rightarrow 2^{+})$
transition probabilities, quadrupole moments $Q(2^{+})$, gyromagnetic
factors $g(2^{+})$ as well as half-lives $T_{1/2}^{2\nu }$ for the $%
0^{+}\rightarrow 0^{+}$ transition and the available experimental data. In
the present work, we study the $\left( \beta ^{-}\beta ^{-}\right) _{0\nu }$
decay for the $0^{+}\rightarrow 0^{+}$ transition in the mass mechanism and
extract limits on effective mass of light as well as heavy neutrinos from
the observed half-lives $T_{1/2}^{0\nu }(0^{+}\rightarrow 0^{+})$ using
nuclear transition matrix elements calculated with the same set of wave
functions. Further, the effect of deformation on the nuclear transition
matrix elements required to study the $\left( \beta ^{-}\beta ^{-}\right)
_{0\nu }$ decay in the mass mechanism is investigated. It is noticed that
the deformation effect on nuclear transition matrix elements is of
approximately same magnitude in $\left( \beta ^{-}\beta ^{-}\right) _{2\nu }$
and $\left( \beta ^{-}\beta ^{-}\right) _{0\nu }$ decay.
\end{abstract}


\maketitle

\section{ INTRODUCTION}

The experimental search for mass and nature of neutrinos through terrestrial
as well as extra-terrestrial approaches has finally confirmed the neutrino
oscillation in atmospheric \cite{suka05,macr04,soud03}, solar \cite
{misc,suka02,snoc02}, reactor \cite{kaml03,kaml04} and accelerator \cite
{k2k05,mino06} neutrino sources and it has been established that neutrinos
have mass. Further, it is generally agreed that oscillations among three
neutrino species are sufficient to explain the atmospheric, solar, reactor
and accelerator neutrino puzzle. The neutrino oscillation experiments
provide us with neutrino mass square differences $\Delta $m$^{2}$, mixing
angles and possible hierarchy in the neutrino mass spectrum. The global 
analysis of atmospheric, solar, reactor (Kam-LAND and CHOOZ) and 
accelarator (K2K and MINOS) neutrino oscillation data with 
2$\sigma $ (3$\sigma $) error provides $\Delta $m$_{12}^{2}$ $=$7.6$%
_{-0.3}^{+0.5}\times $10$^{-5}$ (7.6$_{-0.5}^{+0.7}\times $10$^{-5}$)
eV$^{2}$, sin$^{2}\theta _{12}=0.32_{-0.4}^{+0.5}$ ($0.32_{-0.6}^{+0.8}$) for
solar neutrinos, $\Delta $m$_{23}^{2}$ $=$2.4$\pm 0.3\times $10$^{-3}$ 
(2.4$\pm 0.4\times $10$^{-3}$) eV$^{2}$, sin$^{2}\theta
_{23}=0.50_{-0.12}^{+0.13}$ ($0.50_{-0.16}^{+0.17}$) for atmospheric
neutrinos and $sin^{2}\theta _{13}\leq 0.033$ $(0.050)$ \cite{schw07}. The
oscillation data also suggest that the neutrinos may belong to either normal
hierarchy ($m_{1}<$ $m_{2}<$ $m_{3}$) or inverted hierarchy ($m_{3}<$ $m_{1}$
$<m_{2}$). The data do not exclude the possibility that the mass of the
light neutrino is much larger than $\sqrt{\Delta m_{a}^{2}}$ and this
implies the possible existence of quasi-degenerate neutrino mass spectrum
i.e. $m_{1}\simeq $ $m_{2}\simeq $ $m_{3}$. However, the actual mass of
neutrinos cannot be extracted from these data. On the other hand, the study
of tritium single $\beta $-decay and nuclear double-$\beta$ ($\beta \beta $)
decay together can provide sharpest limits on the mass and nature of
neutrinos.

The nuclear $\beta \beta $ decay is a rare second order
semi-leptonic transition between two even $Z$-even $N$ isobars $_{Z}^{A}$X
and $_{Z\pm 2}^{~~A}$Y involving strangeness conserving charged weak
currents. The implications of present studies about 
nuclear $\beta \beta $ decay,
which is expected to proceed through four experimentally distinguishable
modes, namely the two neutrino double beta $\left( \beta \beta \right)
_{2\nu }$ decay, neutrinoless double beta $(\beta \beta )_{0\nu }$ decay,
single Majoron accompanied $\left( \beta \beta \phi \right) _{0\nu }$decay
and double Majoron accompanied $\left( \beta \beta \phi \phi \right) _{0\nu
} $ decay are far reaching in nature. The half-life of $\left( \beta \beta
\right) _{2\nu }$ decay is a product of accurately known phase space factor
and appropriate nuclear transition matrix element (NTME) $M_{2\nu }$. It has
been already measured in the case of $\left( \beta ^{-}\beta ^{-}\right)
_{2\nu }$ decay for about ten nuclei out of 35 possible candidates \cite
{tret02}. Hence, the values of NTMEs $M_{2\nu }$ can be extracted directly.
Consequently, the validity of different models employed for nuclear
structure calculations can be tested by calculating the $M_{2\nu }$. The $%
\left( \beta \beta \right) _{0\nu }$ decay, which violates the conservation
of lepton number, can occur in a number of gauge theoretical models, namely
GUTs -left-right symmetric models and E(6)-, R$_{p}$-conserving as well as
violating SUSY models, in the scenarios of leptoquark exchange, existence of
heavy sterile neutrino, compositeness and Majoron models. Hence, it is a
convenient tool to test the physics beyond the standard model (SM). In
particular for the question whether the neutrino is a Majorana or Dirac
particle, the $\left( \beta \beta \right) _{0\nu }$ decay is considered to
be the most sensitive way of distinguishing between these two possibilities.
The experimental as well as theoretical developments in the study of nuclear 
$\beta \beta $ decay have been excellently reviewed over the past years
\cite{haxt84,doi85,verg02,faes88,tomo91,boeh92,doi93,moev94,suho98,faes98,
klap86,zube98,fior98,ejir00,elli02,elli04,faes06,avig06,voge06,simk06}.

The $\left( \beta ^{-}\beta ^{-}\right) _{0\nu }$ decay has not been
observed so far and the best observed limit on half-life $T_{1/2}^{0\nu
}>1.9\times 10^{25}$ yr for $^{76}$Ge has been achieved in the
Heidelberg-Moscow experiment \cite{klap01}. Klapdor and his group have
recently reported that $\left( \beta ^{-}\beta ^{-}\right) _{0\nu }$ decay
has been observed in $^{76}$Ge and $T_{1/2}^{0\nu }=1.19\times 10^{25}$ yr 
\cite{klap04}. However, it is felt that the latter result needs independent
verification \cite{aals02,zdes02,elli04}. The aim of all the present
experimental activities is to observe the $\left( \beta ^{-}\beta
^{-}\right) _{0\nu }$ decay. Hence, the models predict half-lives assuming
certain value for the neutrino mass or conversely extract various parameters
from the observed limits of the half-lives of $\left( \beta \beta \right)
_{0\nu }$ decay. The reliability of predictions can be judged a priori only
from the success of a nuclear model in explaining various observed physical
properties of nuclei. The common practice is to calculate the $M_{2\nu }$ to
start with and compare them with the experimentally observed value as the
two decay modes involve the same set of initial and final nuclear wave
functions although the structure of nuclear transition operators are quite
different. Over the past few years, the $\left( \beta \beta \right) _{0\nu }$
decay has been studied mainly in three types of models, namely shell model
and its variants, the quasiparticle random phase approximation (QRPA) and
its extensions and alternative models. In the recent past, the advantages as
well as shortcomings of these models have been excellently discussed by
Suhonen \textit{et al.} \cite{suho98} and Faessler \textit{et al.} \cite
{faes98}.

The structure of nuclei in the mass region $94\leq A\leq 150$ is quite
complex. In the mass region $A\approx 100$, the reduced $B(E2$:$0^{+}\to 2^{+})$ transition probabilities of Zr, Mo, Ru, Pd and Cd isotopes were observed to be as enhanced as in the rare-earth and actinide regions. This mass region offers a nice example of shape transition through sudden onset of deformation at neutron
number $N=60$. The nuclei are soft vibrators for neutron number $N<60$ and
quasi-rotors for $N>60$. The nuclei with neutron number $N=60$ are
transitional nuclei. Similarly, the Te and Xe isotopes in the mass region $%
A\approx 130$ have a vibrational and rotational excitation spectra
respectively. The mass region $A\approx 150$ offers another example of shape
transition i.e. the sudden onset of deformation at neutron number $N=90$.
Nuclei range from spherical to well deformed, with large static quadrupole
moments. In the mass region $94\leq A\leq 150$ of our interest, the
experimental deformation parameter $\beta _{2}$ varies from $0.081\pm 0.016$
to $0.2848\pm 0.0021$ \cite{rama87}. The lowest and highest $\beta _{2}$
values correspond to $^{96}$Zr and $^{150}$Nd respectively. It is well known
that the pairing part of the interaction (\textit{P}) accounts for the
sphericity of nucleus, whereas the quadrupole-quadrupole (\textit{QQ})
interaction increases the collectivity in the nuclear intrinsic wave
functions and makes the nucleus deformed. Hence, it is expected that subtle
interplay of pairing and deformation degrees of freedom will play a crucial
role in reproducing the properties of nuclei in this mass region. All the
nuclei undergoing $\beta \beta $ decay are even-even type, in which the
pairing degrees of freedom play an important role. Moreover, it has been
already conjectured that the deformation can play a crucial role in case of $%
\beta ^{-}\beta ^{-}$ decay of $^{100}$Mo and $^{150}$Nd isotopes \cite
{grif92,suho94}. Hence, it is desirable to have a model which incorporates
the pairing and deformation degrees of freedom on equal footing in its
formalism.

In the light of above discussions, the PHFB model is one of the most natural
choices. However, it is not possible to study the structure of odd-odd
nuclei in the present version of the PHFB model. Hence, the single $\beta $%
-decay rates and the distribution of Gamow-Teller strength can not be
calculated. On the other hand, the study of these processes has implications
in the understanding of the role of the isoscalar part of the proton-neutron
interaction. This is a serious draw back in the present formalism of the
PHFB model. Notwithstanding, the PHFB model in conjunction with the
summation method has been successfully applied to the $\left( \beta
^{-}\beta ^{-}\right) _{2\nu }$ decay of $^{94,96}$Zr, $^{98,100}$Mo, $%
^{104} $Ru and $^{110}$Pd \cite{chan05}, $^{128,130}$Te and $^{150}$Nd \cite
{sing07} isotopes for the $0^{+}\to 0^{+}$ transition not in isolation but
together with other observed nuclear spectroscopic properties, namely yrast
spectra, reduced $B(E2)$ transition probabilities, static quadrupole moments
and $g$-factors. The success of the PHFB model in conjunction with the 
\textit{PPQQ} interaction in explaining $\left( \beta ^{-}\beta ^{-}\right)
_{2\nu }$ decay as well as the above mentioned nuclear properties in the
mass range $94\leq A\leq 150$ has prompted us to apply the same to study the 
$\left( \beta ^{-}\beta ^{-}\right) _{0\nu }$ decay of the same nuclei for
the 0$^{+}\rightarrow $0$^{+}$ transition. Further, the PHFB model using 
the \textit{PPQQ} interaction is a convenient choice to examine the explicit 
role of deformation on NTMEs $M_{2\nu }.$ In the case of 
$\left( \beta ^{-}\beta ^{-}\right) _{2\nu }$ decay, we have observed that 
the deformation plays an important role in the quenching of $M_{2\nu }$ by 
a factor of approximately 2 to 6 \cite{chan05,sing07}. Therefore, we also 
study the variation of NTMEs of $\left( \beta ^{-}\beta ^{-}\right) _{0\nu }$ 
decay vis-a-vis the change in deformation through the changing strength 
of the \textit{QQ} interaction.

Our aim is to extract limits on the effective mass of light as well as heavy
neutrinos from the study of the $\left( \beta ^{-}\beta ^{-}\right) _{0\nu }$
decay of $^{94,96}$Zr, $^{98,100}$Mo, $^{104}$Ru, $^{110}$Pd, $^{128,130}$Te
and $^{150}$Nd isotopes for the 0$^{+}\rightarrow $0$^{+}$ transition. The
present paper is organized as follows. The theoretical formalism to
calculate the half-life of the $\left( \beta \beta \right) _{0\nu }$ decay
mode has been given by Vergados \cite{verg02} and Doi \textit{et al. }\cite
{doi93}. Hence, we briefly outline steps of the above derivations in Sec. II
for clarity in notations used in the present paper following Doi \textit{et
al.} \cite{doi93}. In Sec. III, we present the results and discuss them
vis-a-vis the existing calculations done in other nuclear models. The
deformation effect on NTMEs of $\left( \beta ^{-}\beta ^{-}\right) _{0\nu }$
decay in the mass mechanism is also discussed in the same Section. Finally,
the conclusions are given in Sec. IV.

\section{THEORETICAL FORMALISM}

The effective weak interaction Hamiltonian density for $\beta ^{-}$-decay
due to $W$-boson exchange
restricted to left-handed currents,
 is given by 
\begin{equation}
H_{W}=\frac{G}{\sqrt{2}} j_{L\mu }J_{L}^{\mu \dagger}  +h.c.
\end{equation}
where 
\begin{equation}
\frac{G}{\sqrt{2}} =\frac{g^{2}}{8M_{1}^{2}}\left[ \cos ^{2}\zeta +\left(%
\frac{M_{1}}{M_{2}}\right)^{2}\sin ^{2}\zeta \right] 
\end{equation}
The gauge bosons $W_{L}$ and $W_{R}$ are the superposition of mass
eigenstates $W_{1}$ and $W_{2}$ with masses $M_{1}$ and $M_{2}$ respectively
and mixing angle $\zeta $,
and
$G$ = 1.16637 $ \times $ 10$^{-5}$ GeV$^{-2}.$ 
%
The left 
handed $V-A$ weak leptonic charged current is given
by 
\begin{equation}
j_{L\mu }={\bar{e}\gamma }_{\mu }(1-\gamma _{5})\nu _{eL}
\end{equation}
where 
\begin{equation}
\nu _{eL}=\sum\limits_{i=1}^{2n}\ U_{ei}N_{iL}
\label{eq3}
\end{equation}
Here, $\ N_{i}$\ is a Majorana neutrino field with mass $\ m_{i}$. In Eq. (%
\ref{eq3}), a Dirac neutrino is expressed as a superposition of a pair of
mass degenerate Majorana neutrinos in the most general form. Further, the
mixing parameters are constrained by the following orthonormality
conditions. 
\begin{equation}
\sum\limits_{i=1}^{2n}|U_{ei}|^{2}=\sum\limits_{i=1}^{2n}|V_{ei}|^{2}=1%
\qquad \hbox{and} \qquad \sum\limits_{i=1}^{2n}U_{ei}V_{ei}=0
\end{equation}
The strangeness conserving $V-A$ hadronic currents are given by 
\begin{equation}
J{_{L}^{\mu \dagger }}=g_{V}\overline{u}{\gamma ^{\mu }}(1-\gamma
_{5})d
\end{equation}

\noindent where 
\begin{equation}
g_{V}=\cos \theta _{c}
\label{ckm}
\end{equation}
In Eq. (\ref{ckm}), the $\theta _{c}$ 
is the
Cabibbo-Kobayashi-Maskawa (CKM) mixing angle for the left 
handed $d$ and $s$ quarks. 

In the mass mechanism, the following approximations are taken in deriving
the half-life of $\left( \beta ^{-}\beta ^{-}\right) _{0\nu }$ decay.


\noindent (i) The light neutrino species of mass $m_{i}<10$ MeV and heavy
neutrinos of mass $m_{i}>1$ GeV are considered.

\noindent (ii) In the $0^{+}\rightarrow 0^{+}$ transition, the $s_{1/2}$
waves are considered to describe the final leptonic states.

\noindent (iii) The nonrelativistic impulse approximation is considered for
the hadronic currents.

\noindent (iv) The tensorial terms in the hadronic currents \cite{sim99} are neglected.
While according to Ref. \cite{avi08} they can reduce the matrix elements up to 30\%, in Ref. \cite{suh08} it was reported that its contribution is quite small and can be safely neglected.

\noindent (v) In the case of $\left( \beta ^{-}\beta ^{-}\right) _{0\nu }$
decay, the exchanged neutrinos are virtual particles. Their typical energy $%
\omega =\left( q^{2}+m^{2}\right) ^{1/2}$ is much larger than the typical
excitation energy of the intermediate nuclear states

\begin{equation}
\omega \sim 100\,\mathrm{MeV}\gg E_{N}-E_{I}\sim k^{2}/M\sim 10\,\mathrm{MeV}
\end{equation}
and the variation of $E_{N}$ in the energy denominator can be safely
neglected. Hence, the closure approximation is valid and it is usually used
by replacing the $E_{N}$ by an average $\left\langle E_{N}\right\rangle $.

\noindent (vi) No finite de Broglie wave length correction is considered.
This assumption helps in the calculation of phase space factors.

\noindent (vii) The CP conservation is assumed so that the $U_{ei}$ and $%
V_{ei}$ are both real or purely imaginary depending on the CP parity of the
mass eigenstates of neutrinos $N$ and consequently the effective neutrino
masses $\left\langle m_{\nu }\right\rangle $ and $\left\langle
M_{N}\right\rangle $ for the light and heavy neutrinos are real.

In mass mechanism, the inverse half-life of $\left( \beta ^{-}\beta
^{-}\right) _{0\nu }$ decay in 2n mechanism for the $0^{+}\rightarrow 0^{+}$
transition is given by \cite{verg02,doi93} 
\begin{eqnarray}
\left[ T_{1/2}^{0\nu }(0^{+}\rightarrow 0^{+})\right] ^{-1} &=&\left(\frac
{\left\langle m_{\nu }\right\rangle }{m_{e}}\right)^{2}G_{01}\left(
M_{GT}-M_{F}\right) ^{2}+\left(\frac{m_{p}}{\left\langle M_{N}\right\rangle }%
\right)^{2}G_{01}\left( M_{GTh}-M_{Fh}\right) ^{2}  \nonumber \\
&&+\left(\frac{\left\langle m_{\nu }\right\rangle }{m_{e}}\right)\left(\frac
{m_{p}}{\left\langle M_{N}\right\rangle }\right)G_{01}\left(
M_{GT}-M_{F}\right) \left( M_{GTh}-M_{Fh}\right)  \label{t0n}
\end{eqnarray}

\noindent where 
\begin{eqnarray}
\left\langle m_{\nu }\right\rangle &=&\sum\nolimits_{i}^{\prime
}U_{ei}^{2}m_{i} , ~~~~m_i < 10\hbox{MeV},
\\
\left\langle M_{N}\right\rangle ^{-1} &=&\sum\nolimits_{i}^{\prime \prime
}U_{ei}^{2}m_{i}^{-1}, ~~~~~ m_i > 1 \hbox{GeV}
\end{eqnarray}

\noindent In the closure approximation, NTMEs $M_{\alpha }$ are written as 
\begin{equation}
M_{\alpha }=\sum_{n,m}\left\langle 0_{F}^{+}\left\| O_{\alpha ,nm}\tau
_{n}^{+}\tau _{m}^{+}\right\| 0_{I}^{+}\right\rangle
\end{equation}
where the nuclear transition operators are given by

\begin{eqnarray}
O_{F} &=&\left( \frac{g_{V}}{g_{A}}\right) ^{2}H_{m}(r) \\
O_{GT} &=&\mathbf{\sigma }_{1}\cdot \mathbf{\sigma }_{2}H_{m}(r) \\
O_{Fh} &=&4\pi \left( M_{p}m_{e}\right) ^{-1}\left( \frac{g_{V}}{g_{A}}%
\right) ^{2}\delta \left( \mathbf{r}\right) \\
O_{GTh} &=&4\pi \left( M_{p}m_{e}\right) ^{-1}\mathbf{\sigma }_{1}\cdot 
\mathbf{\sigma }_{2}\delta \left( \mathbf{r}\right)
\end{eqnarray}
The neutrino potentials $H_{m}(r)$ arising due to the exchange of light
neutrinos is defined as 
\begin{eqnarray}
H_{m}\left( r\right) &=&\frac{4\pi R}{\left( 2\pi \right) ^{3}}\int d^{3}q%
\frac{\exp \left( i\mathbf{q}\cdot \mathbf{r}\right) }{\omega \left( \omega +%
\overline{A}\right) }  \nonumber \\
&=&\frac{2R}{\pi r}\int \frac{\sin \left( qr\right) }{\left( q+\overline{A}%
\right) }dq  \nonumber \\
&=&\frac{R}{r}\phi \left( \overline{A}r\right)  \label{hm}
\end{eqnarray}
In going from the first to second line of Eq. (\ref{hm}), the neutrino mass
has been neglected in comparison with the typical neutrino momentum $%
q\approx 200 \, m_{e}$. Further 
\begin{equation}
\overline{A}=\left\langle E_{N}\right\rangle -\frac{1}{2}\left(
E_{I}+E_{F}\right)
\end{equation}
The function $\phi \left( x\right) $ is defined by

\begin{equation}
\phi \left( x\right) =\frac{2}{\pi }\left( \sin x\,ci\left( x\right) -\cos
x\,si\left( x\right) \right)
\end{equation}
where 
\begin{equation}
ci\left( x\right) =-\int_{x}^{\infty }t^{-1}\cos t\,dt \qquad and \qquad
si\left( x\right) =-\int_{x}^{\infty }t^{-1}\sin t\,tdt
\end{equation}
\qquad In the PHFB model, the long range correlations are taken into account
through the configuration mixing while the short range correlations due to
the repulsive hard core is usually absent. In a microscopic picture, the
short range correlations (SRC) arise mainly from the repulsive
nucleon-nucleon potential arising due to the exchange of $\rho $ and $\omega 
$ mesons. Hirsch \textit{et al.} have included the short range effects
through the $\omega $-meson exchange to study the  $\left( \beta ^{-}\beta
^{-}\right) _{0\nu }$ decay in heavy deformed nuclei \cite{jghi95}. The
effect due to SRC has been recently studied by Kortelainen \textit{et al.} \cite{kort07} and Simkovic \textit{et al.} \cite{simk08}
in the unitary correlation operator method (UCOM). This SRC
effect can also be incorporated phenomenologically by using the Jastrow
type of correlation given by the prescription

\begin{equation}
\left\langle j_{1}^{\pi }j_{2}^{\pi }J\left| O\right| j_{1}^{\nu }j_{2}^{\nu
}J^{^{\prime }}\right\rangle \rightarrow \left\langle j_{1}^{\pi }j_{2}^{\pi
}J\left| f\:O\:f\right| j_{1}^{\nu }j_{2}^{\nu }J^{^{\prime }}\right\rangle
\end{equation}
where

\begin{equation}
f(r)=1-e^{-ar^{2}}(1-br^{2})\label{jast}
\end{equation}
with $a$ = 1.1 fm$^{-2}$ and $b$ = 0.68 fm$^{-2}$. 
However, it is not clear which approach is the best \cite{simk08}. Therefore, 
we include the short range correlation by using the Jastrow correlations 
given by Eq. (\ref{jast})
in the present work. Wu and co-workers \cite{wu85} derived the 
effective transition operator $\widehat{f}O\widehat{f}$
for the $\left( \beta ^{-}\beta ^{-}\right) _{0\nu }$ decay of $^{48}$Ca
using Reid and Paris potentials. The calculations show that
phenomenologically determined $f(r)$ has strong two nucleon correlations. 
The consideration of the effect due to SRC requires the inclusion of finite
size of the nucleon and is taken into account by the replacements

\begin{equation}
g_{V}\rightarrow g_{V}\left( \frac{\Lambda ^{2}}{\Lambda ^{2}+k^{2}}\right)
^{2}\qquad and \qquad g_{A}\rightarrow g_{A}\left( \frac{\Lambda ^{2}}{%
\Lambda ^{2}+k^{2}}\right) ^{2}
\end{equation}
with $\Lambda $ = 850 MeV. Including finite size effect, the neutrino
potential for the exchange of heavy neutrinos $U_{0}(r,\Lambda )$ is written
as 
\begin{eqnarray}
U_{0}(r,\Lambda ) &=&\frac{R}{\left( 2\pi \right) ^{3}}\int d\mathbf{k}e^{i%
\mathbf{k}\cdot \mathbf{r}}\left( \frac{\Lambda ^{2}}{\Lambda ^{2}+k^{2}}%
\right) ^{4}  \nonumber \\
&=&\frac{R}{2\pi ^{2}r}\int \sin \left( qr\right) \left( \frac{\Lambda ^{2}}{%
\Lambda ^{2}+k^{2}}\right) ^{4}qdq \\
&=&\frac{\Lambda ^{3}R}{64\pi }e^{-\Lambda r}\left[ 1+\Lambda r+\frac{1}{3}%
\left( \Lambda r\right) ^{2}\right]
\end{eqnarray}

The expression to calculate the NTMEs $M_{\alpha }$ of $\left( \beta
^{-}\beta ^{-}\right) _{0\nu }$ decay for the $0^{+}\rightarrow 0^{+}$
transition in the PHFB model is obtained as follows. In the PHFB model, a
state with good angular momentum \textbf{J} is obtained from the axially
symmetric HFB intrinsic state $|\Phi _{0}\rangle $ with $K=0$ through the
following relation using the standard projection technique \cite{onis66}

\begin{equation}
|\Psi _{00}^{J}\rangle =\left[ \frac{(2J+1)}{{8\pi ^{2}}}\right] \int
D_{00}^{J}(\Omega )R(\Omega )|\Phi _{0}\rangle d\Omega
\end{equation}
where $\ R(\Omega )$\ and $\ D_{MK}^{J}(\Omega )$\ are the rotation operator
and the rotation matrix respectively. The axially symmetric HFB intrinsic
state ${|\Phi _{0}\rangle }$ can be written as 
\begin{equation}
{|\Phi _{0}\rangle }=\prod\limits_{im}(u_{im}+v_{im}b_{im}^{\dagger }b_{i%
\bar{m}}^{\dagger })|0\rangle
\end{equation}
where the creation operators $\ b_{im}^{\dagger }$\ and $\ b_{i\bar{m}%
}^{\dagger }$\ are defined as 
\begin{equation}
b{_{im}^{\dagger }}=\sum\limits_{\alpha }C_{i\alpha ,m}a_{\alpha m}^{\dagger
}\quad and \quad b_{i\bar{m}}^{\dagger }=\sum\limits_{\alpha
}(-1)^{l+j-m}C_{i\alpha ,m}a_{\alpha ,-m}^{\dagger }
\end{equation}
\ The amplitudes $\ (u_{im},v_{im})$\ and the expansion coefficient $\
C_{ij,m}$ are obtained from the HFB calculations.

Employing the HFB wave functions, one obtains the following expression for
NTMEs $M_{\alpha }$ of $\left( \beta ^{-}\beta ^{-}\right) _{0\nu }$ decay 
\cite{dixi02}. 
\begin{eqnarray}
M_{\alpha } &=&\left[ n^{Ji=0}n^{J_{f}=0}\right] ^{-1/2}\int\limits_{0}^{\pi
}n_{(Z,N),(Z+2,N-2)}(\theta )\frac{1}{4}\sum\limits_{\alpha \beta \gamma
\delta }\langle \alpha \beta \left| O_{\alpha }\right| \gamma \delta \rangle
\nonumber \\
&&\times \sum\limits_{\varepsilon \eta }\frac{\left( f_{Z+2,N-2}^{(\pi
)*}\right) _{\varepsilon \beta }}{\left[ \left( 1+F_{Z,N}^{(\pi )}(\theta
)f_{Z+2,N-2}^{(\pi )*}\right) \right] _{\varepsilon \alpha }^{-1}}\frac{%
\left( F_{Z,N}^{(\nu )*}\right) _{\eta \delta }}{\left[ \left(
1+F_{Z,N}^{(\nu )}(\theta )f_{Z+2,N-2}^{(\nu )*}\right) \right] _{\gamma
\eta }^{-1}}sin\theta d\theta  \label{ntm}
\end{eqnarray}
where 
\begin{equation}
n^{J}=\int\limits_{0}^{\pi }\left[ det\left( 1+F^{(\pi )}f^{(\pi )^{\dagger
}}\right) \right] ^{1/2}\left[ det\left( 1+F^{(\nu )}f^{(\nu )^{\dagger
}}\right) \right] ^{1/2}d_{00}^{J}(\theta )sin(\theta )d\theta
\end{equation}
and 
\begin{equation}
n_{(Z,N),(Z+2,N-2)}{(\theta )}=\left[ det\left( 1+F_{Z,N}^{(\nu
)}f_{Z+2,N-2}^{(\nu )^{\dagger }}\right) \right] ^{1/2}\times \left[
det\left( 1+F_{Z,N}^{(\pi )}f_{Z+2,N-2}^{(\pi )^{\dagger }}\right) \right]
^{1/2}
\end{equation}
The $\pi (\nu )$\ represents the proton (neutron) of nuclei involved in the $%
\left( \beta \beta \right) _{0\nu }$ decay process. The matrix $%
F_{Z,N}(\theta )$ and $f_{Z,N}$ are given by 
\begin{eqnarray}
F_{Z,N}(\theta ) &=&\sum\limits_{m_{\alpha }^{\prime }m_{\beta }^{\prime
}}d_{m_{\alpha },m_{\alpha }^{\prime }}^{j_{\alpha }}(\theta )d_{m_{\beta
},m_{\beta }^{\prime }}^{j_{\beta }}(\theta )f_{j_{\alpha }m_{\alpha
}^{\prime },j_{\beta }m_{\beta }^{\prime }} \\
f_{Z,N} &=&\sum\limits_{i}C_{ij_{\alpha },m_{\alpha }}C_{ij_{\beta
},m_{\beta }}\left( v_{im_{\alpha }}/u_{im_{\alpha }}\right) \delta
_{m_{\alpha },-m_{\beta }}
\end{eqnarray}
The calculation of required NTMEs $M_{\alpha }$ are performed in the
following manner. In the first step, matrices $f^{(\pi ,\nu )}$ and$\
F^{(\pi ,\nu )}(\theta )$ are setup for the nuclei involved in the $\left(
\beta \beta \right) _{0\nu }$ decay making use of 20 Gaussian quadrature
points in the range (0,$\pi $). Finally, the required NTMEs can be calculated in a straight forward manner  using the Eq. (\ref{ntm}).

It is worth to underline the fact that the building blocks for constructing 
the HFB states are spherical harmonic oscillator states, which depend only 
on the mass number $A$  and there is no dependence on $N$ and $Z$. For this reason the overlap expressions given above are exact in the present context, and the ``two vacua" problem found in QRPA calculations \cite{civ87} is not present.

\section{RESULTS AND DISCUSSIONS}

In the present work, we calculate the NTMEs $M_{F}$, $M_{GT}$, $M_{Fh}$ and $%
M_{GTh}$ for $^{94,96}$Zr, $^{98,100}$Mo, $^{104}$Ru, $^{110}$Pd, $%
^{128,130}$Te and $^{150}$Nd isotopes to study the $0^{+}\to 0^{+}$
transition of $\left( \beta ^{-}\beta ^{-}\right) _{0\nu }$ decay in the
mass mechanism using a set of HFB wave functions, the reliability of which
was tested by obtaining an over all agreement between theoretically
calculated results for the yrast spectra, reduced $B(E2$:$0^{+}\rightarrow
2^{+})$ transition probabilities, static quadrupole moments $Q(2^{+}),$ $g$%
-factors $g(2^{+})$ and NTMEs $M_{2\nu }$ as well as half-lives $%
T_{1/2}^{2\nu }$ of $\left( \beta ^{-}\beta ^{-}\right) _{2\nu }$ decay and
the available experimental data \cite{chan05,sing07}. However, we briefly
discuss in the following the model space, single particle energies (SPE's)
and parameters of two-body interactions used to generate the HFB wave
functions for convenience.

We treat the doubly even nucleus $^{76}$Sr ($N=Z=38$) as an inert core in
case of $^{94,96}$Zr, $^{94,96,98,100}$Mo, $^{98,100,104}$Ru, $^{104,110}$Pd
and $^{110}$Cd nuclei, with the valence space spanned by $1p_{1/2},$ $%
2s_{1/2,}$ $1d_{3/2},$ $1d_{5/2},$ $0g_{7/2},$ $0g_{9/2}$ and $0h_{11/2}$
orbits for protons and neutrons. The $1p_{1/2}$ orbit has been included in
the valence space to examine the role of the $Z=40$ proton core vis-a-vis
the onset of deformation in the highly neutron rich isotopes. The set of
single particle energies (SPE's) used were (in MeV) $\varepsilon
(1p_{1/2})=-0.8$, $\varepsilon (0g_{9/2})=0.0$, $\varepsilon (1d_{5/2})=5.4$%
, $\varepsilon (2s_{1/2})=6.4$, $\varepsilon (1d_{3/2})=7.9$, $\varepsilon
(0g_{7/2})=8.4$ and $\varepsilon (0h_{11/2})=8.6$ for proton and neutron.

In case of $^{128,130}$Te, $^{128,130}$Xe, $^{150}$Nd and $^{150}$Sm nuclei,
we treated the doubly even nucleus $^{100}$Sn ($N=Z=50$) as an inert core with
the valence space spanned by $2s_{1/2}$, $1d_{3/2}$, $1d_{5/2}$, $1f_{7/2},$ 
$0g_{7/2}$, $0h_{9/2}$ and $0h_{11/2}$ orbits for protons and neutrons. The
change of model space was forced upon due to the following reason. In the model space used for mass
region $A \approx 100$, the number of neutrons increase to about 40 for nuclei
occurring in the mass region $A \approx 130$. With the increase 
in neutron number,
the yrast energy spectra gets compressed due to increase in the attractive
part of effective two-body interaction. The set of single particle energies
(SPE's) used were in MeV: $\varepsilon (1d_{5/2})=0.0$, $\varepsilon
(2s_{1/2})=1.4$, $\varepsilon (1d_{3/2})=2.0$, $\varepsilon (0g_{7/2})=4.0$, 
$\varepsilon (0h_{11/2})=6.5$ (4.8 for $^{150}$Nd and $^{150}$Sm), $%
\varepsilon (1f_{7/2})=12.0$ (11.5 for $^{150}$Nd and $^{150}$Sm), $%
\varepsilon (0h_{9/2})=12.5$ (12.0 for $^{150}$Nd and $^{150}$Sm) for proton
and neutron.

The HFB\ wave functions were generated using an effective Hamiltonian with 
\textit{PPQQ} type \cite{bara68} of two-body interaction. Explicitly, the
Hamiltonian can be written as 
\begin{equation}
H=H_{sp}+V(P)+\zeta _{qq}V(QQ)
\end{equation}
where $H_{sp}$ denotes the single particle Hamiltonian. The $V(P)$ and
$V(QQ)$ represent the pairing and quadrupole-quadrupole part of the
effective two-body interaction. The $\ \zeta _{qq}$ is an arbitrary
parameter and the final results were obtained by setting the $\ \zeta _{qq}$
= 1. The purpose of introducing $\zeta _{qq}$ was to study the role of
deformation by varying the strength of \textit{QQ} interaction. The
strengths of the pairing interaction was fixed through the relation $%
G_{p}=30/A$ MeV and $G_{n}=20/A$ MeV, which were the same as used by Heestand 
\textit{et al.} \cite{hees69} to explain the experimental \textit{g}(2$^{+}$%
) data of some even-even Ge, Se, Mo, Ru, Pd, Cd and Te isotopes in Greiner's
collective model \cite{grei66}. For $^{96}$Zr, we have used $G_{n}$ = 22/%
\textit{A} MeV. The strengths of the pairing interaction fixed for $%
^{128,130}$Te, $^{128,130}$Xe, $^{150}$Nd and $^{150}$Sm were $G_{p}=35/A$
MeV and $G_{n}=35/A$ MeV.

The strengths of the like particle components of the \textit{QQ} interaction
were taken as: $\chi _{pp}=\chi _{nn}=0.0105$ MeV \textit{b}$^{-4}$, where 
\textit{b} is the oscillator parameter. The strength of proton-neutron (\textit{%
pn}) component of the \textit{QQ} interaction $\chi _{pn}$ was varied so as
to obtain the spectra of considered nuclei, namely $^{94,96}$Zr, $%
^{94,96,98,100}$Mo, $^{98,100,104}$Ru, $^{104,110}$Pd, $^{110}$Cd, $%
^{128,130}$Te, $^{128,130}$Xe, $^{150}$Nd and $^{150}$Sm in optimum
agreement with the experimental results. The theoretical spectra has been
taken to be the optimum one if the excitation energy of the $2^{+}$
state $E_{2^{+}}$ is reproduced as closely as possible to the experimental
value. The prescribed set of parameters for the strength of \textit{QQ}
interaction 
are in general different for the initial and final nuclei. They have 
been tabulated in the previous works describing the 
$\left(\beta^{-}\beta^{-}\right)_{2\nu}$ decay \cite{chan05,sing07}, and are
%
consistent with those of Arima suggested on the basis of an
empirical analysis of effective two-body interaction \cite{arim81}.

\subsection{Results of $\ \left( \beta ^{-}\beta ^{-}\right) _{0\nu }$ decay}

The phase space factors $G_{01}$ of $\left( \beta ^{-}\beta ^{-}\right)
_{0\nu }$ decay for the $0^{+}\rightarrow 0^{+}$ transition have been
calculated by Doi \textit{et al.} \cite{doi85} and Boehm and Vogel \cite
{boeh92} for $g_{A}=1.25$. We use the later in our calculations after reevaluating them for $g_{A}=1.254$. The average energy denominator $\overline{A}$ is
calculated  using Haxton's prescription given by $\overline{A}%
=1.12A^{1/2} $ MeV \cite{haxt84}. In Table I, we compile all the required
NTMEs $M_{F}$, $M_{GT}$, $M_{Fh}$ and $M_{GTh}\,$for $^{94,96}$Zr, $%
^{98,100} $Mo, $^{104}$Ru, $^{110}$Pd, $^{128,130}$Te and $^{150}$Nd nuclei.
The four NTMEs are calculated in the approximation of point nucleons, point
nucleons plus Jastrow type of SRC, finite size of nucleons with dipole form
factor and finite size plus SRC. In the case of point nucleons, the NTMEs $%
M_{F}$ and $M_{GT}$ are calculated for $\overline{A}$ and $\overline{A}/2$
in the energy denominator$.$ It is observed that the NTMEs $M_{F}$ and $%
M_{GT}$ change by 7.5-10.3 $\%$ for $\overline{A}/2$ in comparison to $%
\overline{A}$ in the energy denominator. Hence, the dependence of NTMEs on
average excitation energy $\overline{A}$ is small and the closure
approximation is quite good in the case of $\left( \beta ^{-}\beta
^{-}\right) _{0\nu }$ decay as expected. In the approximation of light
neutrinos, the NTMEs $M_{F}$ and $M_{GT}$ are reduced  by  17-22\% and 11.6-14.4\% for point nucleon plus SRC and finite size of nucleons respectively.
Finally the NTMEs change by 20.8-26.5\% with finite size plus SRC. In case
of heavy neutrinos, the $M_{Fh}$ and $M_{GTh}$ get reduced  by 34.5-39.3\% and
65.4-69.5\% with the inclusion of finite size and finite size plus SRC.

\subsection{Limits on effective mass of light and heavy neutrinos}

In Table II, we compile the calculated NTMEs $M_{F}$, $M_{GT}$, $M_{Fh}$ and $%
M_{GTh}$ of the above mentioned isotopes for the $0^{+}\to 0^{+}$ transition
in the PHFB as well as other\ models. Further, we also present the results
of shell model \cite{caur07}, QRPA as well as RQRPA calculations by 
Rodin {\it et al.} \cite{rodi07}, Simkovic {\it et al.} \cite{simk08} 
and pseudo SU(3) \cite{jghi95} for light neutrinos.
Besides using different model space, single particle energies and effective
two-body interactions, the value of $g_{A}$ used in these calculation is
different. To be specific, $g_{A}=1.0$ \cite{jghi95}, $g_{A}=1.25$ \cite
{simk01,caur07,rodi07} and $g_{A}=1.254$ \cite{muto89,tomo91,hirs96,simk08}. 
The value of $g_{A}$ is not available for calculation of 
Pantis \textit{et al}. \cite{pant94,pant96}. Recently, Simkovic {\it et al.} 
\cite{simk08} have evaluated $M^{(0\nu)}$ using both QRPA and 
RQRPA taking an average of 24 calculations. 
Their smallest and largest NTME $M^{(0\nu)}$, calculated in 
the QRPA(RQRPA) with Jastrow type of correlations,
are included in Table II, and for this reason two values are given for each nuclei associated with Ref. \cite{simk08}.  
%
The spreads in NTMEs 
$M^{(0\nu)}$($M^{(0\nu)}_{N}$) of light(heavy) neutrinos for $^{96}$Zr, $^{100}$Mo, 
$^{128,130}$Te and $^{150}$Nd isotopes are 6.7(8.5), 127.2(40.1), 5.4(7.0),
2.8(4.2) and 3.9(9.3) respectively. Excluding the result of FQRPA \cite{pant96},
the same spreads turn out to be 2.4(2.1), 4.2(5.8), 2.9(5.9), 2.5(4.2) and
3.9(9.3) respectively. 
The convergence of the NTMEs calculated in different models to a factor smaller than 4 is encouraging. And it should be taken into account that some calculations do not include deformation effects, and that the size of contribution form the tensorial terms is still being debated and deserves further studies.

We extract the limits on effective light neutrino mass 
$\left\langle m_{\nu }\right\rangle $ and heavy neutrino mass $\left\langle
M_{N}\right\rangle $ using the phase space factors evaluated for $%
g_{A}=1.254 $. However, the NTMEs $M_{F}$ and $M_{Fh}$ need to be evaluated
for the same $g_{A}$. We tabulate the extracted limits on effective light
neutrino mass $\left\langle m_{\nu }\right\rangle $ and heavy neutrino mass $%
\left\langle M_{N}\right\rangle $ in the same Table II using experimental
limits on half-lives $T_{1/2}^{0\nu }$ given below. 

\begin{center}
\begin{tabular}{llllll}
$T_{1/2}^{0\nu }>1.9\times 10^{19}$ yr & ($^{94}$Zr) & \cite{arno99} &
$~~~~~~T_{1/2}^{0\nu }>6.0\times 10^{17}$ yr & ($^{110}$Pd) & \cite{wint52} \\
$T_{1/2}^{0\nu }>1.0\times 10^{21}$ yr & ($^{96}$Zr) & \cite{arno99} &
$~~~~~~T_{1/2}^{0\nu }>1.1\times 10^{23}$ yr & ($^{128}$Te)& \cite{arna03} \\
$T_{1/2}^{0\nu }>1.0\times 10^{14}$ yr & ($^{98}$Mo) & \cite{frem52} &
$~~~~~~T_{1/2}^{0\nu }>3.0\times 10^{24}$ yr & ($^{130}$Te) & \cite{arna08} \\
$T_{1/2}^{0\nu }>4.6\times 10^{23}$ yr & ($^{100}$Mo) & \cite{arno05} &
$~~~~~~T_{1/2}^{0\nu }>1.2\times 10^{21}$ yr & ($^{150}$Nd)& \cite{desi97}
\end{tabular}
\end{center}

No experimental half-life limit on $\left( \beta ^{-}\beta ^{-}\right)
_{0\nu }$ decay is available in the case of $^{104}$Ru isotope. In case of $%
^{94}$Zr, $^{98}$Mo and $^{110}$Pd isotopes, the observed experimental
half-life limits are small and no theoretical calculation has been performed
so far. The extracted upper and lower bounds on $\left\langle m_{\nu
}\right\rangle $ and $\left\langle M_{N}\right\rangle $ from the observed
half-life limits \textit{T}$_{1/2}^{0\nu }$ using NTMEs calculated in the
PHFB model are $1.40\times 10^{3}$ eV, $3.60\times 10^{6}$ eV, $1.44\times
10^{3}$ eV and $1.05\times 10^{4}$ GeV, 3.8 GeV and $1.01\times 10^{4}$ GeV
for $^{94}$Zr, $^{98}$Mo and $^{110}$Pd respectively$.$

In the case of $^{96}$Zr isotope, the theoretically calculated 
$\left| M^{\left( 0\nu \right) }\right| $ in
the PHFB model and RQRPA \cite{rodi07} are close. The $\left|
M^{\left( 0\nu \right) }\right| $ calculated in the FQRPA \cite{pant96} and
QPRA \cite{rodi07} are smaller than the presently calculated value by
factors of 4.1 and 1.3 respectively while the $\left| M^{\left( 0\nu \right)
}\right| $ calculated in the QRPA \cite{pant96} is larger by a factor of
1.7. In case of heavy neutrinos, the only available $\left| M_{N}^{\left(
0\nu \right) }\right| $ is due to Pantis \textit{\textit{et al.}} \cite
{pant96}. Using the NTMEs calculated in the PHFB model, the extracted 
limits on light and heavy neutrino mass from the
observed half-life limit $T_{1/2}^{0\nu }>1.0\times 10^{21}$ yr \cite{arno99}
are $\left\langle m_{\nu }\right\rangle <$ 45.66 eV and $\left\langle
M_{N}\right\rangle >$ 3.4$\times $10$^{5}$ GeV respectively.

The $\left( \beta ^{-}\beta ^{-}\right) _{0\nu }$ decay of $^{100}$Mo is a
theoretically well studied case. The NTME $\left| M^{\left( 0\nu \right)
}\right| $ calculated in the PHFB model lies in the range of $\left|
M^{\left( 0\nu \right) }\right| $ calculated in QRPA \cite{rodi07} and close
to RQRPA \cite{simk01}. The NTMEs $\left| M^{\left( 0\nu \right) }\right| $
calculated in other nuclear models \cite{muto89,tomo91,pant94,pant96} are
small in comparison to the NTMEs calculated in the PHFB model. There are six
matrix elements available in case of heavy neutrinos. The NTMEs $\left|
M_{N}^{\left( 0\nu \right) }\right| $ calculated in the QRPA by Tomoda \cite
{tomo91} and Hirsch \textit{\textit{et al}.} \cite{hirs96} are larger by a
factor of 1.6 and 3.3 respectively than the present work. All other matrix
elements $\left| M_{N}^{\left( 0\nu \right) }\right| $are smaller than 
the presently calculated value. 
The extracted upper and lower bounds on light and heavy
neutrino masses $\left\langle m_{\nu }\right\rangle $ and $\left\langle
M_{N}\right\rangle $ from the observed half-life limit $T_{1/2}^{0\nu
}>4.6\times 10^{23}$ yr \cite{arno05} in the PHFB model are 1.08 eV and $%
1.32\times 10^{7}$ GeV respectively.

The NTME $\left| M^{\left( 0\nu \right) }\right| $ calculated in the PHFB
model and QRPA \cite{tomo91} for $^{128}$Te isotope 
are close and larger by a factor
of 1.8 than the calculated NTME in FQRPA \cite{pant96}. All other NTMEs $%
\left| M^{\left( 0\nu \right) }\right| $ are large in comparison to the
result of PHFB model. In case of heavy neutrinos, 
the NTMEs $\left| M_{N}^{\left( 0\nu \right) }\right| $ 
have been calculated by Tomoda \cite{tomo91}, Hirsch 
\textit{\textit{et al.}} \cite{hirs96} and Pantis \textit{\textit{et al.}} 
\cite{pant96}. The $\left| M_{N}^{\left( 0\nu \right) }\right| $ calculated
in our PHFB model is larger than the calculated NTME in FQRPA \cite{pant96}
by a factor of 1.2 and smaller than results of Tomoda \cite{tomo91}, Hirsch 
\textit{\textit{et al }}\cite{hirs96} and Pantis \textit{\textit{et al.}} 
\cite{pant96}. The extracted neutrino mass limits using our NTMEs from the
observed half-life limit $T_{1/2}^{0\nu }>1.1\times 10^{23}$ yr \cite{arna03}
are $\left\langle m_{\nu }\right\rangle <$ 22.05 eV and $\left\langle
M_{N}\right\rangle > 6.84 \times 10^{5}$ GeV respectively.

In the case of $^{130}$Te isotope, the $\left| M^{\left( 0\nu \right) }\right|$
calculated in the PHFB model, SM \cite{caur07} and QRPA \cite{pant96} are
close. It is smaller than the $\left| M^{\left( 0\nu \right) }\right| $
calculated in QRPA and its extensions \cite{rodi07,muto89,pant96}. The
presently calculated $\left| M^{\left( 0\nu \right) }\right| $ is larger by
a factor of 1.3 and 1.5 than the calculated values in QPPA \cite{tomo91} and
FQRPA \cite{pant96}. In case of heavy neutrinos, our calculated $\left|
M_{N}^{\left( 0\nu \right) }\right| $ in the PHFB model is smaller than the
available NTMEs $\left| M_{N}^{\left( 0\nu \right) }\right| $ calculated in
other models$.$ The extracted neutrino mass limits from the observed
half-life limit $T_{1/2}^{0\nu }>3.0\times 10^{24}$ yr \cite{arna08} are $%
\left\langle m_{\nu }\right\rangle $ $<$ 0.63 eV and $\left\langle
M_{N}\right\rangle \ >$ 2.18$\times $10$^{7}$ GeV respectively using NTMEs
of the present work.

The NTMEs $\left| M^{\left( 0\nu \right) }\right| $ have been calculated by
Tomoda \cite{tomo91}, Hirsch \textit{\textit{et al.}} \cite{jghi95}, Rodin 
\textit{et al.} \cite{rodi07} and Muto \textit{\textit{et al.}} \cite{muto89}
for $^{150}$Nd isotope. The $\left| M^{\left( 0\nu \right) }\right| $ 
calculated in
the PHFB model and pseudo SU(3] \cite{jghi95} are close and smaller by a
factor of 3.8 than the $\left| M^{\left( 0\nu \right) }\right| $ calculated
in the QRPA by Muto \textit{et al.} \cite{muto89}. The $\left| M^{\left(
0\nu \right) }\right| $ calculated by Rodin \textit{et al.} \cite{rodi07} 
in QRPA and RQRPA are large in comparison to our calculated results.
However, the deformation effect has not been taken into account in the
calculation of Rodin $et$ $al$. \cite{rodi07}. Arguably, the results based on the
spherical QRPA are larger than ours in the present case as expected.
In case of heavy neutrinos, our calculated $\left| M_{N}^{\left( 0\nu
\right) }\right| $ in the PHFB model is smaller than the reported NTMEs in
the QRPA \cite{tomo91,hirs96}. Using the NTMEs of our PHFB model, the
extracted upper and lower bounds on $\left\langle m_{\nu }\right\rangle $
and $\left\langle M_{N}\right\rangle $ from the observed half-life limit $%
T_{1/2}^{0\nu }>1.2\times 10^{21}$ yr \cite{desi97} are 19.85 eV and $%
6.79\times 10^{5}$ GeV respectively.

\subsection{Quadrupolar correlations and deformation}

To understand the role of 
quadrupolar correlations
on the NTMEs $M_{\alpha }$ $(\alpha
=F,GT,Fh,GTh)$ of $\left( \beta ^{-}\beta ^{-}\right) _{0\nu }$ decay for
the 0$^{+}\rightarrow 0^{+}$ transition, we investigate the variation of the 
$M_{\alpha }$ by changing the strength parameter $\zeta _{qq}$ of \textit{QQ}
part of the effective two-body interaction. In Table III, we present the
NTMEs $M_{\alpha }$ of $^{94,96}$Zr, $^{98,100}$Mo, $^{104}$Ru, $^{110}$Pd, $%
^{128,130}$Te and $^{150}$Nd nuclei for different $\zeta _{qq}$. In general,
it is observed that NTMEs $M_{\alpha }$ remain initially almost constant and
then start decreasing as the strength parameter $\zeta _{qq}$ approaches the
physical value 1. There are of course few minor observed anomalies which can
not be explained at the present stage. 
This reduction in the magnitude of NTMEs from thier maximum values obtained 
with pure pairing has a clear analogy with their behavior in the presence of 
particle-particle $1^+$ proton-neutron correlations \cite{vog86} which has been 
widely discussed in the literature (see Ref. \cite{avi08} and references therein).

To quantify the effect of quadrupolar interaction on $M_{\alpha }$,
which is closely related with the nuclear deformation,
 we define a quantity 
$D_{\alpha }$ as the ratio of $M_{\alpha }$ 
calculated with pure pairing 
($\zeta_{qq}=0$) and 
and with full presence of the quadrupolar interaction
($\zeta _{qq}=1$). Explicitly, the $D_{\alpha}$ is given by 
\begin{equation}
D_{\alpha }=\frac{M_{\alpha }(\zeta _{qq}=0)}{M_{\alpha }(\zeta _{qq}=1)}
\end{equation}
The values of $D_{\alpha }$ are tabulated in Table IV for the considered
nuclei. These values of $D_{\alpha }$ suggest that the NTMEs $M_{\alpha }$
due to light and heavy neutrino exchange are quenched by factors of 1.92 to
6.25 and 1.92 to 6.27 respectively in the mass region $94\leq A\leq 150$ due
to 
the presence of quadrupolar correlations.
For comparison, we also give the 
ratio $D_{2\nu }$ \cite{chan05,sing07} 
associated with the same effect in the $(\beta^{-}\beta^{-})_{2\nu}$ matrix elements,
in the last row of the same table, which also change by
almost same amount due to the 
quadrupolar interaction. 
Hence, it is clear that the deformation effects are important in case of $\left( \beta ^{-}\beta
^{-}\right) _{2\nu }$ as well as $\left( \beta ^{-}\beta ^{-}\right) _{0\nu
} $ decay so far as the nuclear structure aspect of $\beta ^{-}\beta
^{-}$ decay is concerned.

A suppression of the double beta matrix element with respect to the spherical case has been reported when the parent and daughter nuclei have different deformations \cite{alva06}. To investigate this effect
we present in Fig. 1 the NTMEs for $^{150}$Nd as a function of the difference in the deformation parameter $\beta_2$ between the parent and daughter nuclei. The NTMEs are calculated keeping the deformation for parent nuclei fixed at 
$\zeta_{qq}$=1 and the deformation of daughter nuclei are varied by taking $\zeta_{qq}=$0.0 to 1.5. 

\begin{figure}
\scalebox{0.5}{\includegraphics{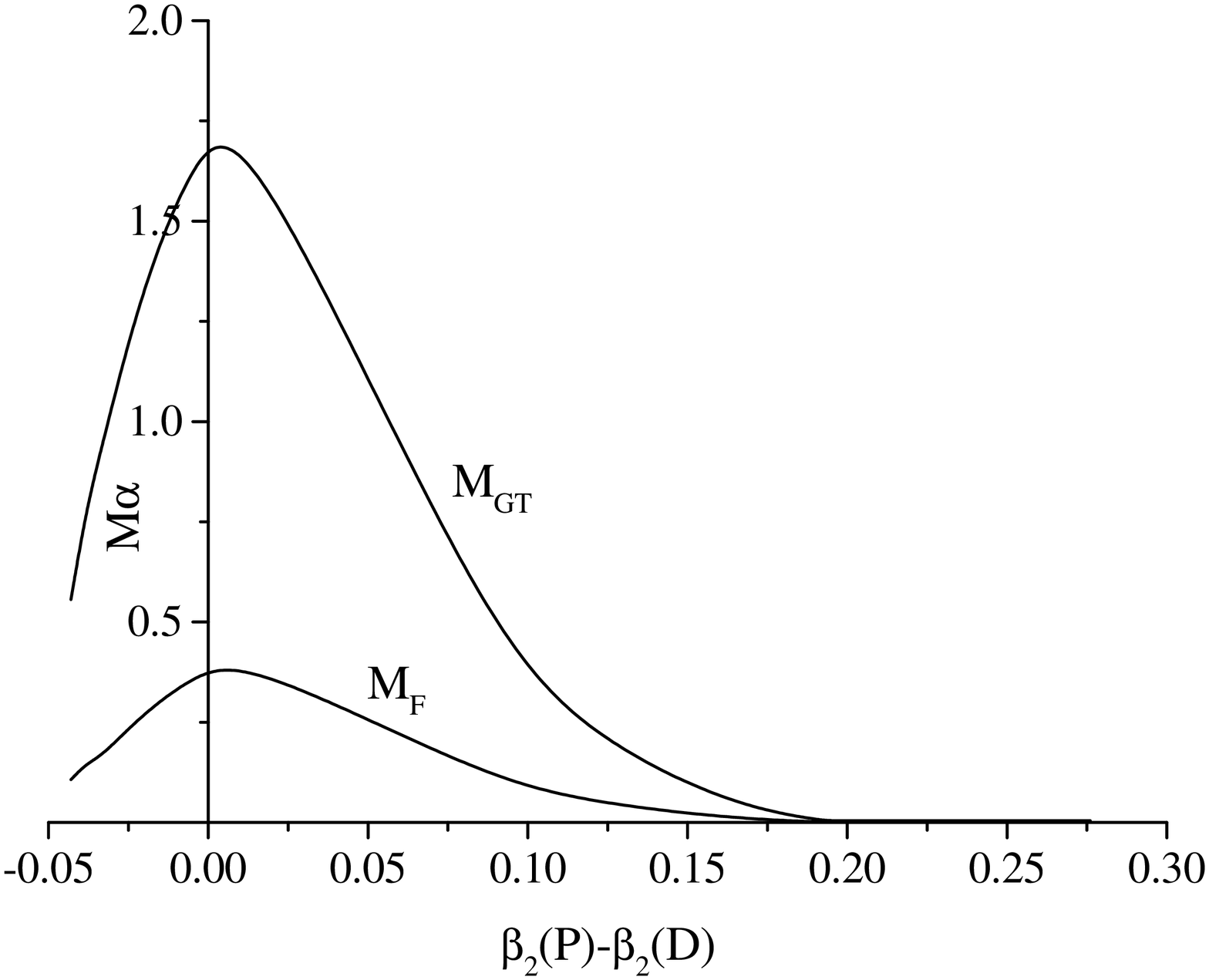}}
\caption{\label{fig1} $\beta\beta0\nu$ NTME for $^{150}$Nd as a function of the difference in the deformation parameter $\beta_2$ between the parent and daughter nuclei.}
\vspace{-1.0pc}
\end{figure}

It is clear that when the deformations of parent and daughter nuclei are
similar the NTMEs have a well defined maximum, and when the absolute value
of the difference in deformation increases the NTMEs are strongly reduced,
in full agreement with the results reported in \cite{alva06}.
In the case of $^{150}$%
Nd$\rightarrow ^{150}$Sm transition, the large difference in the observed
deformation parameters $\beta _{2}$ of parent and daughter isotopes suggests
that the reduction of NTMEs is substantial. In the present calculation, the
magnitude of reduction could be underestimated due to the difference in the
theoretically calculated \cite{sing07} and experimentally observed 
parametric values of deformations.

\section{CONCLUSIONS}

The required NTMEs to study $\left( \beta ^{-}\beta ^{-}\right) _{0\nu }$ \
decay of $^{94,96}$Zr, $^{98,100}$Mo, $^{104}$Ru, $^{110}$Pd, $^{128,130}$Te
and $^{150}$Nd isotopes in mass mechanism are calculated in the PHFB model.
It is observed that the NTMEs have a weak dependence on the average
excitation energy $\overline{A}$ of intermediate nucleus and the closure
approximation is quite valid, as expected. Further, we extract limits on
gauge theoretical parameters, namely $\left\langle m_{\nu }\right\rangle $
and $\left\langle M_{N}\right\rangle $ from the available limits on
experimental half-lives $T_{1/2}^{0\nu }$ using NTMEs calculated in the PHFB
as well as other models. For the best candidates $^{100}$Mo and $^{130}$Te
nuclei, the PHFB, SM\ \cite{caur07} and QRPA \cite{pant96} results are
close. Finally, the calculated values of $D_{\alpha }$
suggest that the NTMEs $M_{\alpha }$ due to light and heavy neutrino
exchange are quenched by factors of 1.92 to 6.25 and 1.92 to 6.27
respectively in the mass region $94\leq A\leq 150$ due to the deformation. 
In $^{150}$Nd the difference in deformation between the parent and daughter nuclei is small and adds no extra suppression.
We conclude that nuclear structure effects are also
important for the $\left( \beta ^{-}\beta ^{-}\right) _{0\nu }$ \ decay.

This work was partially supported by DST, India vide sanction No.
SR/S2/HEP-13/2006, by Conacyt-M\'{e}xico and DGAPA-UNAM.

\begin{table}
\caption{NTMEs $M_{F}$, $M_{GT}$, $M_{Fh}$ and $M_{GTh}$
for the $\left( \beta ^{-}\beta ^{-}\right) _{0\nu }$ \ decay of $^{94,96}$%
Zr, $^{98,100}$Mo, $^{104}$Ru, $^{110}$Pd, $^{128,130}$Te and $^{150}$Nd
isotopes.}

\begin{tabular}{p{0.7in}p{0.25in}p{0.8in}p{0.8in}p{0.8in}p{0.8in}p{0.8in}}
\hline\hline
Nuclei &NTMEs & \multicolumn{2}{c}
{~~~~~~~~Point} & ~~~~~~~~~~~~Point & ~~~~~~~~Extened
& ~~~~~~~~Extended \\ 
\multicolumn{1}{l}{} & \multicolumn{1}{l}{} & ~~~~~~~~~~~~$\overline{A}$ 
&~~~~~~~~~~~$\overline{A}/2$ & \multicolumn{1}{r}{+SRC} & \multicolumn{1}{r}{} & 
\multicolumn{1}{r}{+SRC} \\ \hline
\multicolumn{1}{l}{} & \multicolumn{1}{l}{} & \multicolumn{1}{l}{} & 
\multicolumn{1}{l}{} & \multicolumn{1}{l}{} & \multicolumn{1}{l}{} & 
\multicolumn{1}{l}{} \\ 
\multicolumn{1}{l}{$^{94}$Zr} & \multicolumn{1}{l}{$M_{F}$} & 
\multicolumn{1}{r}{0.4634} & \multicolumn{1}{r}{0.4983} & \multicolumn{1}{r}{
0.3640} & \multicolumn{1}{r}{0.3975} & \multicolumn{1}{r}{0.3433} \\ 
\multicolumn{1}{l}{} & \multicolumn{1}{l}{$M_{GT}$} & \multicolumn{1}{r}{
-2.2557} & \multicolumn{1}{r}{-2.4381} & \multicolumn{1}{r}{-1.7871} & 
\multicolumn{1}{r}{-1.9470} & \multicolumn{1}{r}{-1.6912} \\ 
\multicolumn{1}{l}{} & \multicolumn{1}{l}{$M_{Fh}$} & \multicolumn{1}{r}{
35.1339} & \multicolumn{1}{r}{} & \multicolumn{1}{r}{0} & \multicolumn{1}{r}{
21.8092} & \multicolumn{1}{r}{11.1290} \\ 
\multicolumn{1}{l}{} & \multicolumn{1}{l}{$M_{GTh}$} & \multicolumn{1}{r}{
-165.746} & \multicolumn{1}{r}{} & \multicolumn{1}{r}{0} & 
\multicolumn{1}{r}{-101.736} & \multicolumn{1}{r}{-51.5522} \\ 
\multicolumn{1}{l}{} & \multicolumn{1}{l}{} & \multicolumn{1}{r}{} & 
\multicolumn{1}{r}{} & \multicolumn{1}{r}{} & \multicolumn{1}{r}{} & 
\multicolumn{1}{r}{} \\ 
\multicolumn{1}{l}{$^{96}$Zr} & \multicolumn{1}{l}{$M_{F}$} & 
\multicolumn{1}{r}{0.3494} & \multicolumn{1}{r}{0.3757} & \multicolumn{1}{r}{
0.2734} & \multicolumn{1}{r}{0.2991} & \multicolumn{1}{r}{0.2576} \\ 
\multicolumn{1}{l}{} & \multicolumn{1}{l}{$M_{GT}$} & \multicolumn{1}{r}{
-1.6264} & \multicolumn{1}{r}{-1.7614} & \multicolumn{1}{r}{-1.2681} & 
\multicolumn{1}{r}{-1.3923} & \multicolumn{1}{r}{-1.1958} \\ 
\multicolumn{1}{l}{} & \multicolumn{1}{l}{$M_{Fh}$} & \multicolumn{1}{r}{
26.7691} & \multicolumn{1}{r}{} & \multicolumn{1}{r}{0} & \multicolumn{1}{r}{
16.6927} & \multicolumn{1}{r}{8.5271} \\ 
\multicolumn{1}{l}{} & \multicolumn{1}{l}{$M_{GTh}$} & \multicolumn{1}{r}{
-126.284} & \multicolumn{1}{r}{} & \multicolumn{1}{r}{0} & 
\multicolumn{1}{r}{-76.6661} & \multicolumn{1}{r}{-38.4867} \\ 
\multicolumn{1}{l}{} & \multicolumn{1}{l}{} & \multicolumn{1}{r}{} & 
\multicolumn{1}{r}{} & \multicolumn{1}{r}{} & \multicolumn{1}{r}{} & 
\multicolumn{1}{r}{} \\ 
\multicolumn{1}{l}{$^{98}$Mo} & \multicolumn{1}{l}{$M_{F}$} & 
\multicolumn{1}{r}{0.7951} & \multicolumn{1}{r}{0.8639} & \multicolumn{1}{r}{
0.6474} & \multicolumn{1}{r}{0.6960} & \multicolumn{1}{r}{0.6159} \\ 
\multicolumn{1}{l}{} & \multicolumn{1}{l}{$M_{GT}$} & \multicolumn{1}{r}{
-3.5921} & \multicolumn{1}{r}{-3.9033} & \multicolumn{1}{r}{-2.8955} & 
\multicolumn{1}{r}{-3.1293} & \multicolumn{1}{r}{-2.7502} \\ 
\multicolumn{1}{l}{} & \multicolumn{1}{l}{$M_{Fh}$} & \multicolumn{1}{r}{
52.1823} & \multicolumn{1}{r}{} & \multicolumn{1}{r}{0} & \multicolumn{1}{r}{
33.2343} & \multicolumn{1}{r}{17.2313} \\ 
\multicolumn{1}{l}{} & \multicolumn{1}{l}{$M_{GTh}$} & \multicolumn{1}{r}{
-246.172} & \multicolumn{1}{r}{} & \multicolumn{1}{r}{0} & 
\multicolumn{1}{r}{-153.939} & \multicolumn{1}{r}{-78.8982} \\ 
\multicolumn{1}{l}{} & \multicolumn{1}{l}{} & \multicolumn{1}{r}{} & 
\multicolumn{1}{r}{} & \multicolumn{1}{r}{} & \multicolumn{1}{r}{} & 
\multicolumn{1}{r}{} \\ 
\multicolumn{1}{l}{$^{100}$Mo} & \multicolumn{1}{l}{$M_{F}$} & 
\multicolumn{1}{r}{0.7849} & \multicolumn{1}{r}{0.8476} & \multicolumn{1}{r}{
0.6339} & \multicolumn{1}{r}{0.6831} & \multicolumn{1}{r}{0.6013} \\ 
\multicolumn{1}{l}{} & \multicolumn{1}{l}{$M_{GT}$} & \multicolumn{1}{r}{
-3.5121} & \multicolumn{1}{r}{-3.7994} & \multicolumn{1}{r}{-2.8007} & 
\multicolumn{1}{r}{-3.0375} & \multicolumn{1}{r}{-2.6505} \\ 
\multicolumn{1}{l}{} & \multicolumn{1}{l}{$M_{Fh}$} & \multicolumn{1}{r}{
53.1825} & \multicolumn{1}{r}{} & \multicolumn{1}{r}{0} & \multicolumn{1}{r}{
34.3112} & \multicolumn{1}{r}{17.9096} \\ 
\multicolumn{1}{l}{} & \multicolumn{1}{l}{$M_{GTh}$} & \multicolumn{1}{r}{
-250.891} & \multicolumn{1}{r}{} & \multicolumn{1}{r}{0} & 
\multicolumn{1}{r}{-158.446} & \multicolumn{1}{r}{-81.6304} \\ 
\multicolumn{1}{l}{} & \multicolumn{1}{l}{} & \multicolumn{1}{r}{} & 
\multicolumn{1}{r}{} & \multicolumn{1}{r}{} & \multicolumn{1}{r}{} & 
\multicolumn{1}{r}{} \\ 
\multicolumn{1}{l}{$^{104}$Ru} & \multicolumn{1}{l}{$M_{F}$} & 
\multicolumn{1}{r}{0.6073} & \multicolumn{1}{r}{0.6582} & \multicolumn{1}{r}{
0.4934} & \multicolumn{1}{r}{0.5300} & \multicolumn{1}{r}{0.4684} \\ 
\multicolumn{1}{l}{} & \multicolumn{1}{l}{$M_{GT}$} & \multicolumn{1}{r}{
-2.5306} & \multicolumn{1}{r}{-2.7285} & \multicolumn{1}{r}{-1.9938} & 
\multicolumn{1}{r}{-2.1704} & \multicolumn{1}{r}{-1.8787} \\ 
\multicolumn{1}{l}{} & \multicolumn{1}{l}{$M_{Fh}$} & \multicolumn{1}{r}{
40.0647} & \multicolumn{1}{r}{} & \multicolumn{1}{r}{0} & \multicolumn{1}{r}{
26.2356} & \multicolumn{1}{r}{13.8055} \\ 
\multicolumn{1}{l}{} & \multicolumn{1}{l}{$M_{GTh}$} & \multicolumn{1}{r}{
-189.007} & \multicolumn{1}{r}{} & \multicolumn{1}{r}{0} & 
\multicolumn{1}{r}{-120.874} & \multicolumn{1}{r}{-62.7040} \\ 
\multicolumn{1}{l}{} & \multicolumn{1}{l}{} & \multicolumn{1}{r}{} & 
\multicolumn{1}{r}{} & \multicolumn{1}{r}{} & \multicolumn{1}{r}{} & 
\multicolumn{1}{r}{} \\ 
\multicolumn{1}{l}{$^{110}$Pd} & \multicolumn{1}{l}{$M_{F}$} & 
\multicolumn{1}{r}{0.9020} & \multicolumn{1}{r}{0.9802} & \multicolumn{1}{r}{
0.7260} & \multicolumn{1}{r}{0.7835} & \multicolumn{1}{r}{0.6880} \\ 
\multicolumn{1}{l}{} & \multicolumn{1}{l}{$M_{GT}$} & \multicolumn{1}{r}{
-4.1660} & \multicolumn{1}{r}{-4.5419} & \multicolumn{1}{r}{-3.3365} & 
\multicolumn{1}{r}{-3.6125} & \multicolumn{1}{r}{-3.1611} \\ 
\multicolumn{1}{l}{} & \multicolumn{1}{l}{$M_{Fh}$} & \multicolumn{1}{r}{
61.9271} & \multicolumn{1}{r}{} & \multicolumn{1}{r}{0} & \multicolumn{1}{r}{
40.0043} & \multicolumn{1}{r}{20.8651} \\ 
\multicolumn{1}{l}{} & \multicolumn{1}{l}{$M_{GTh}$} & \multicolumn{1}{r}{
-292.144} & \multicolumn{1}{r}{} & \multicolumn{1}{r}{0} & 
\multicolumn{1}{r}{-185.378} & \multicolumn{1}{r}{-95.6358} \\ 
\multicolumn{1}{l}{} & \multicolumn{1}{l}{} & \multicolumn{1}{r}{} & 
\multicolumn{1}{r}{} & \multicolumn{1}{r}{} & \multicolumn{1}{r}{} & 
\multicolumn{1}{r}{} \\ 
\multicolumn{1}{l}{$^{128}$Te} & \multicolumn{1}{l}{$M_{F}$} & 
\multicolumn{1}{r}{0.4063} & \multicolumn{1}{r}{0.4460} & \multicolumn{1}{r}{
0.3273} & \multicolumn{1}{r}{0.3536} & \multicolumn{1}{r}{0.3107} \\ 
\multicolumn{1}{l}{} & \multicolumn{1}{l}{$M_{GT}$} & \multicolumn{1}{r}{
-1.7644} & \multicolumn{1}{r}{-1.9237} & \multicolumn{1}{r}{-1.3913} & 
\multicolumn{1}{r}{-1.5172} & \multicolumn{1}{r}{-1.3142} \\ 
\multicolumn{1}{l}{} & \multicolumn{1}{l}{$M_{Fh}$} & \multicolumn{1}{r}{
28.0755} & \multicolumn{1}{r}{} & \multicolumn{1}{r}{0} & \multicolumn{1}{r}{
17.7056} & \multicolumn{1}{r}{9.1133} \\ 
\multicolumn{1}{l}{} & \multicolumn{1}{l}{$M_{GTh}$} & \multicolumn{1}{r}{
-132.448} & \multicolumn{1}{r}{} & \multicolumn{1}{r}{0} & 
\multicolumn{1}{r}{-82.3872} & \multicolumn{1}{r}{-42.0190} \\ 
\multicolumn{1}{l}{} & \multicolumn{1}{l}{} & \multicolumn{1}{r}{} & 
\multicolumn{1}{r}{} & \multicolumn{1}{r}{} & \multicolumn{1}{r}{} & 
\multicolumn{1}{r}{} \\ 
\multicolumn{1}{l}{$^{130}$Te} & \multicolumn{1}{l}{$M_{F}$} & 
\multicolumn{1}{r}{0.5252} & \multicolumn{1}{r}{0.5787} & \multicolumn{1}{r}{
0.4306} & \multicolumn{1}{r}{0.4616} & \multicolumn{1}{r}{0.4105} \\ 
\multicolumn{1}{l}{} & \multicolumn{1}{l}{$M_{GT}$} & \multicolumn{1}{r}{
-2.3468} & \multicolumn{1}{r}{-2.5723} & \multicolumn{1}{r}{-1.9001} & 
\multicolumn{1}{r}{-2.0472} & \multicolumn{1}{r}{-1.8055} \\ 
\multicolumn{1}{l}{} & \multicolumn{1}{l}{$M_{Fh}$} & \multicolumn{1}{r}{
33.6801} & \multicolumn{1}{r}{} & \multicolumn{1}{r}{0} & \multicolumn{1}{r}{
21.4747} & \multicolumn{1}{r}{11.1405} \\ 
\multicolumn{1}{l}{} & \multicolumn{1}{l}{$M_{GTh}$} & \multicolumn{1}{r}{
-158.887} & \multicolumn{1}{r}{} & \multicolumn{1}{r}{0} & 
\multicolumn{1}{r}{-100.999} & \multicolumn{1}{r}{-52.2720} \\ 
\multicolumn{1}{l}{} & \multicolumn{1}{l}{} & \multicolumn{1}{r}{} & 
\multicolumn{1}{r}{} & \multicolumn{1}{r}{} & \multicolumn{1}{r}{} & 
\multicolumn{1}{r}{} \\ 
\multicolumn{1}{l}{$^{150}$Nd} & \multicolumn{1}{l}{$M_{F}$} & 
\multicolumn{1}{r}{0.3837} & \multicolumn{1}{r}{0.4231} & \multicolumn{1}{r}{
0.3182} & \multicolumn{1}{r}{0.3391} & \multicolumn{1}{r}{0.3039} \\ 
\multicolumn{1}{l}{} & \multicolumn{1}{l}{$M_{GT}$} & \multicolumn{1}{r}{
-1.6881} & \multicolumn{1}{r}{-1.8606} & \multicolumn{1}{r}{-1.3786} & 
\multicolumn{1}{r}{-1.4787} & \multicolumn{1}{r}{-1.3119} \\ 
\multicolumn{1}{l}{} & \multicolumn{1}{l}{$M_{Fh}$} & \multicolumn{1}{r}{
23.3705} & \multicolumn{1}{r}{} & \multicolumn{1}{r}{0} & \multicolumn{1}{r}{
15.3104} & \multicolumn{1}{r}{8.0695} \\ 
\multicolumn{1}{l}{} & \multicolumn{1}{l}{$M_{GTh}$} & \multicolumn{1}{r}{
-110.251} & \multicolumn{1}{r}{} & \multicolumn{1}{r}{0} & 
\multicolumn{1}{r}{-71.4285} & \multicolumn{1}{r}{-37.3676} \\ 
\multicolumn{1}{l}{} & \multicolumn{1}{l}{} & \multicolumn{1}{r}{} & 
\multicolumn{1}{r}{} & \multicolumn{1}{r}{} & \multicolumn{1}{r}{} & 
\multicolumn{1}{r}{} \\ \hline\hline
\end{tabular}
\end{table}

\begin{table}
\caption{Upper and lower bounds on light and haevy
neutrino masses $<m_{\nu }>$ (eV) and $<M_{N}>$ (GeV) respectively for the $%
\left( \beta ^{-}\beta ^{-}\right) _{0\nu }$ \ decay of $^{94,96}$Zr, $%
^{98,100}$Mo, $^{110}$Pd, $^{128,130}$Te and $^{150}$Nd isotopes in
different nuclear models.}

\begin{tabular}{p{0.6in}p{0.6in}p{0.4in}p{0.6in}p{0.6in}p{0.6in}p{0.6in}p{0.6in}p{0.6in}p{0.6in}p{0.6in}}
\hline\hline
\multicolumn{1}{l}{\small Nuclei} & \multicolumn{1}{l}{\small Model} & 
\multicolumn{1}{l}{\small Ref.} & \multicolumn{1}{r}{$M_{F}$} & 
\multicolumn{1}{r}{$M_{GT}$} & \multicolumn{1}{r}{$\left| M^{\left( 0\nu
\right) }\right| $} & \multicolumn{1}{r}{$<m_{\nu }>$} & \multicolumn{1}{r}{$%
M_{Fh}$} & \multicolumn{1}{r}{$M_{GTh}$} & \multicolumn{1}{r}{$\left|
M_{N}^{\left( 0\nu \right) }\right| $} & \multicolumn{1}{r}{$<M_{N}>$} \\ 
\hline
&  &  &  &  &  &  &  &  &  &  \\ 
\multicolumn{1}{l}{$^{94}${\small Zr}} & \multicolumn{1}{l}{\small PHFB} & 
\multicolumn{1}{l}{\small *} & \multicolumn{1}{r}{\small 0.3433} & 
\multicolumn{1}{r}{\small -1.6912} & \multicolumn{1}{r}{\small 2.0345} & 
\multicolumn{1}{r}{{\small 1.40}$\times ${\small 10}$^{3}$} & 
\multicolumn{1}{r}{\small 11.1290} & \multicolumn{1}{r}{\small -51.5522} & 
\multicolumn{1}{r}{\small 62.6812} & \multicolumn{1}{r}{{\small 1.05}$\times 
${\small 10}$^{4}$} \\ 
\multicolumn{1}{l}{} & \multicolumn{1}{l}{} & \multicolumn{1}{l}{} & 
\multicolumn{1}{r}{} & \multicolumn{1}{r}{} & \multicolumn{1}{r}{} & 
\multicolumn{1}{r}{} & \multicolumn{1}{r}{} & \multicolumn{1}{r}{} & 
\multicolumn{1}{r}{} & \multicolumn{1}{r}{} \\ 
\multicolumn{1}{l}{$^{96}${\small Zr}} & \multicolumn{1}{l}{\small PHFB} & 
\multicolumn{1}{l}{\small *} & \multicolumn{1}{r}{\small 0.2576} & 
\multicolumn{1}{r}{\small -1.1958} & \multicolumn{1}{r}{\small 1.4534} & 
\multicolumn{1}{r}{\small 45.66} & \multicolumn{1}{r}{\small 8.5271} & 
\multicolumn{1}{r}{\small -38.4867} & \multicolumn{1}{r}{\small 47.0138} & 
\multicolumn{1}{r}{{\small 3.40}$\times ${\small 10}$^{5}$} \\ 
\multicolumn{1}{l}{} & \multicolumn{1}{l}{\small QRPA} & \multicolumn{1}{l}{%
{\small \cite{pant96}}} & \multicolumn{1}{r}{\small -0.312} & 
\multicolumn{1}{r}{\small 2.097} & \multicolumn{1}{r}{\small 2.409} & 
\multicolumn{1}{r}{\small 27.55} & \multicolumn{1}{r}{} & \multicolumn{1}{r}{
} & \multicolumn{1}{r}{\small 99.062} & \multicolumn{1}{r}{{\small 7.16}$%
\times ${\small 10}$^{5}$} \\ 
\multicolumn{1}{l}{} & \multicolumn{1}{l}{\small FQRPA} & \multicolumn{1}{l}{%
{\small \cite{pant96}}} & \multicolumn{1}{r}{\small 0.639} & 
\multicolumn{1}{r}{\small 0.280} & \multicolumn{1}{r}{\small 0.359} & 
\multicolumn{1}{r}{\small 184.8} & \multicolumn{1}{r}{} & \multicolumn{1}{r}{
} & \multicolumn{1}{r}{\small 11.659} & \multicolumn{1}{r}{{\small 8.42}$%
\times ${\small 10}$^{4}$} \\ 
\multicolumn{1}{l}{} & \multicolumn{1}{l}{\small RQRPA} & \multicolumn{1}{l}{%
{\small \cite{rodi07}}} & \multicolumn{1}{r}{} & \multicolumn{1}{r}{} & 
\multicolumn{1}{r}{{\small 1.20}$_{-0.14}^{+0.14}$} & \multicolumn{1}{r}%
{\small 55.30} & \multicolumn{1}{r}{} & \multicolumn{1}{r}{} & 
\multicolumn{1}{r}{} & \multicolumn{1}{r}{} \\ 
\multicolumn{1}{l}{} & \multicolumn{1}{l}{\small QRPA} & \multicolumn{1}{l}{%
{\small \cite{rodi07}}} & \multicolumn{1}{r}{} & \multicolumn{1}{r}{} & 
\multicolumn{1}{r}{{\small 1.12}$_{-0.03}^{+0.03}$} & \multicolumn{1}{r}%
{\small 59.25} & \multicolumn{1}{r}{} & \multicolumn{1}{r}{} & 
\multicolumn{1}{r}{} & \multicolumn{1}{r}{} \\ 
\multicolumn{1}{l}{} & \multicolumn{1}{l}{\small RQRPA} & \multicolumn{1}{l}{%
{\small \cite{simk08}}} & \multicolumn{1}{r}{} & \multicolumn{1}{r}{} & 
\multicolumn{1}{r}{\small 1.01} & \multicolumn{1}{r}%
{\small 65.70} & \multicolumn{1}{r}{} & \multicolumn{1}{r}{} & 
\multicolumn{1}{r}{} & \multicolumn{1}{r}{} \\ 
\multicolumn{1}{l}{} & \multicolumn{1}{l}{\small RQRPA} & \multicolumn{1}{l}{%
{\small \cite{simk08}}} & \multicolumn{1}{r}{} & \multicolumn{1}{r}{} & 
\multicolumn{1}{r}{\small 1.31} & \multicolumn{1}{r}%
{\small 50.65} & \multicolumn{1}{r}{} & \multicolumn{1}{r}{} & 
\multicolumn{1}{r}{} & \multicolumn{1}{r}{} \\ 
\multicolumn{1}{l}{} & \multicolumn{1}{l}{\small QRPA} & \multicolumn{1}{l}{%
{\small \cite{simk08}}} & \multicolumn{1}{r}{} & \multicolumn{1}{r}{} & 
\multicolumn{1}{r}{\small 1.34} & \multicolumn{1}{r}%
{\small 49.52} & \multicolumn{1}{r}{} & \multicolumn{1}{r}{} & 
\multicolumn{1}{r}{} & \multicolumn{1}{r}{} \\ 
\multicolumn{1}{l}{} & \multicolumn{1}{l}{\small QRPA} & \multicolumn{1}{l}{%
{\small \cite{simk08}}} & \multicolumn{1}{r}{} & \multicolumn{1}{r}{} & 
\multicolumn{1}{r}{\small 1.79} & \multicolumn{1}{r}%
{\small 37.07} & \multicolumn{1}{r}{} & \multicolumn{1}{r}{} & 
\multicolumn{1}{r}{} & \multicolumn{1}{r}{} \\ 
\multicolumn{1}{l}{} & \multicolumn{1}{l}{} & \multicolumn{1}{l}{} & 
\multicolumn{1}{r}{} & \multicolumn{1}{r}{} & \multicolumn{1}{r}{} & 
\multicolumn{1}{r}{} & \multicolumn{1}{r}{} & \multicolumn{1}{r}{} & 
\multicolumn{1}{r}{} & \multicolumn{1}{r}{} \\ 
\multicolumn{1}{l}{$^{98}${\small Mo}} & \multicolumn{1}{l}{\small PHFB} & 
\multicolumn{1}{l}{\small *} & \multicolumn{1}{r}{\small 0.6159} & 
\multicolumn{1}{r}{\small -2.7502} & \multicolumn{1}{r}{\small 3.3661} & 
\multicolumn{1}{r}{{\small 3.60}$\times ${\small 10}$^{6}$} & 
\multicolumn{1}{r}{\small 17.2313} & \multicolumn{1}{r}{\small -78.8982} & 
\multicolumn{1}{r}{\small 96.1295} & \multicolumn{1}{r}{\small 3.8} \\ 
\multicolumn{1}{l}{} & \multicolumn{1}{l}{} & \multicolumn{1}{l}{} & 
\multicolumn{1}{r}{} & \multicolumn{1}{r}{} & \multicolumn{1}{r}{} & 
\multicolumn{1}{r}{} & \multicolumn{1}{r}{} & \multicolumn{1}{r}{} & 
\multicolumn{1}{r}{} & \multicolumn{1}{r}{} \\ 
\multicolumn{1}{l}{$^{100}${\small Mo}} & \multicolumn{1}{l}{\small PHFB} & 
\multicolumn{1}{l}{\small *} & \multicolumn{1}{r}{\small 0.6013} & 
\multicolumn{1}{r}{\small -2.6505} & \multicolumn{1}{r}{\small 3.2518} & 
\multicolumn{1}{r}{\small 1.08} & \multicolumn{1}{r}{\small 17.9096} & 
\multicolumn{1}{r}{\small -81.6304} & \multicolumn{1}{r}{\small 99.5400} & 
\multicolumn{1}{r}{{\small 1.32}$\times ${\small 10}$^{7}$} \\ 
\multicolumn{1}{l}{} & \multicolumn{1}{l}{\small QRPA} & \multicolumn{1}{l}{%
{\small \cite{muto89,hirs96}}} & \multicolumn{1}{r}{\small -1.356} & 
\multicolumn{1}{r}{\small 0.763} & \multicolumn{1}{r}{\small 2.119} & 
\multicolumn{1}{r}{\small 1.65} & \multicolumn{1}{r}{\small -64.0} & 
\multicolumn{1}{r}{\small 269.0} & \multicolumn{1}{r}{\small 333.0} & 
\multicolumn{1}{r}{{\small 4.56}$\times ${\small 10}$^{7}$} \\ 
\multicolumn{1}{l}{} & \multicolumn{1}{l}{\small QRPA} & \multicolumn{1}{l}{%
{\small \cite{tomo91}}} & \multicolumn{1}{r}{\small -0.588} & 
\multicolumn{1}{r}{\small 1.76} & \multicolumn{1}{r}{\small 2.348} & 
\multicolumn{1}{r}{\small 1.49} & \multicolumn{1}{r}{\small -29.141} & 
\multicolumn{1}{r}{\small 126.819} & \multicolumn{1}{r}{\small 155.960} & 
\multicolumn{1}{r}{{\small 2.14}$\times ${\small 10}$^{7}$} \\ 
\multicolumn{1}{l}{} & \multicolumn{1}{l}{\small QRPA} & \multicolumn{1}{l}{%
{\small \cite{pant94}}} & \multicolumn{1}{r}{\small 0.272} & 
\multicolumn{1}{r}{\small 2.86} & \multicolumn{1}{r}{\small 2.588} & 
\multicolumn{1}{r}{\small 1.35} & \multicolumn{1}{r}{} & \multicolumn{1}{r}{}
& \multicolumn{1}{r}{\small 56.914} & \multicolumn{1}{r}{{\small 7.80}$%
\times ${\small 10}$^{6}$} \\ 
\multicolumn{1}{l}{} & \multicolumn{1}{l}{\small QRPA} & \multicolumn{1}{l}{%
{\small \cite{pant96}}} & \multicolumn{1}{r}{\small -0.471} & 
\multicolumn{1}{r}{\small 0.615} & \multicolumn{1}{r}{\small 1.086} & 
\multicolumn{1}{r}{\small 3.22} & \multicolumn{1}{r}{} & \multicolumn{1}{r}{}
& \multicolumn{1}{r}{\small 76.752} & \multicolumn{1}{r}{{\small 1.05}$%
\times ${\small 10}$^{7}$} \\ 
\multicolumn{1}{l}{} & \multicolumn{1}{l}{\small FQRPA} & \multicolumn{1}{l}{%
{\small \cite{pant96}}} & \multicolumn{1}{r}{\small -0.548} & 
\multicolumn{1}{r}{\small -0.584} & \multicolumn{1}{r}{\small 0.036} & 
\multicolumn{1}{r}{\small 97.16} & \multicolumn{1}{r}{} & \multicolumn{1}{r}{
} & \multicolumn{1}{r}{\small 8.304} & \multicolumn{1}{r}{{\small 1.14}$%
\times ${\small 10}$^{6}$} \\ 
\multicolumn{1}{l}{} & \multicolumn{1}{l}{\small RQRPA} & \multicolumn{1}{l}{%
{\small \cite{simk01}}} & \multicolumn{1}{r}{\small -0.819} & 
\multicolumn{1}{r}{\small 2.620} & \multicolumn{1}{r}{\small 3.439} & 
\multicolumn{1}{r}{\small 1.02} & \multicolumn{1}{r}{\small -28.352} & 
\multicolumn{1}{r}{\small 34.200} & \multicolumn{1}{r}{\small 62.552} & 
\multicolumn{1}{r}{{\small 8.57}$\times ${\small 10}$^{6}$} \\ 
\multicolumn{1}{l}{} & \multicolumn{1}{l}{\small RQRPA} & \multicolumn{1}{l}{%
{\small \cite{rodi07}}} & \multicolumn{1}{r}{} & \multicolumn{1}{r}{} & 
\multicolumn{1}{r}{{\small 2.78}$_{-0.19}^{+0.19}$} & \multicolumn{1}{r}%
{\small 1.26} & \multicolumn{1}{r}{} & \multicolumn{1}{r}{} & 
\multicolumn{1}{r}{} & \multicolumn{1}{r}{} \\ 
\multicolumn{1}{l}{} & \multicolumn{1}{l}{\small QRPA} & \multicolumn{1}{l}{%
{\small \cite{rodi07}}} & \multicolumn{1}{r}{} & \multicolumn{1}{r}{} & 
\multicolumn{1}{r}{{\small 3.34}$_{-0.19}^{+0.19}$} & \multicolumn{1}{r}%
{\small 1.05} & \multicolumn{1}{r}{} & \multicolumn{1}{r}{} & 
\multicolumn{1}{r}{} & \multicolumn{1}{r}{} \\ 
\multicolumn{1}{l}{} & \multicolumn{1}{l}{\small RQRPA} & \multicolumn{1}{l}{%
{\small \cite{simk08}}} & \multicolumn{1}{r}{} & \multicolumn{1}{r}{} & 
\multicolumn{1}{r}{\small 2.22} & \multicolumn{1}{r}%
{\small 1.58} & \multicolumn{1}{r}{} & \multicolumn{1}{r}{} & 
\multicolumn{1}{r}{} & \multicolumn{1}{r}{} \\ 
\multicolumn{1}{l}{} & \multicolumn{1}{l}{\small RQRPA} & \multicolumn{1}{l}{%
{\small \cite{simk08}}} & \multicolumn{1}{r}{} & \multicolumn{1}{r}{} & 
\multicolumn{1}{r}{\small 2.77} & \multicolumn{1}{r}%
{\small 1.26} & \multicolumn{1}{r}{} & \multicolumn{1}{r}{} & 
\multicolumn{1}{r}{} & \multicolumn{1}{r}{} \\ 
\multicolumn{1}{l}{} & \multicolumn{1}{l}{\small QRPA} & \multicolumn{1}{l}{%
{\small \cite{simk08}}} & \multicolumn{1}{r}{} & \multicolumn{1}{r}{} & 
\multicolumn{1}{r}{\small 3.53} & \multicolumn{1}{r}%
{\small 0.99} & \multicolumn{1}{r}{} & \multicolumn{1}{r}{} & 
\multicolumn{1}{r}{} & \multicolumn{1}{r}{} \\ 
\multicolumn{1}{l}{} & \multicolumn{1}{l}{\small QRPA} & \multicolumn{1}{l}{%
{\small \cite{simk08}}} & \multicolumn{1}{r}{} & \multicolumn{1}{r}{} & 
\multicolumn{1}{r}{\small 4.58} & \multicolumn{1}{r}%
{\small 0.76} & \multicolumn{1}{r}{} & \multicolumn{1}{r}{} & 
\multicolumn{1}{r}{} & \multicolumn{1}{r}{} \\ 
\multicolumn{1}{l}{} & \multicolumn{1}{l}{} & \multicolumn{1}{l}{} & 
\multicolumn{1}{r}{} & \multicolumn{1}{r}{} & \multicolumn{1}{r}{} & 
\multicolumn{1}{r}{} & \multicolumn{1}{r}{} & \multicolumn{1}{r}{} & 
\multicolumn{1}{r}{} & \multicolumn{1}{r}{} \\ 
\multicolumn{1}{l}{$^{110}${\small Pd}} & \multicolumn{1}{l}{\small PHFB} & 
\multicolumn{1}{l}{\small *} & \multicolumn{1}{r}{\small 0.6880} & 
\multicolumn{1}{r}{\small -3.1611} & \multicolumn{1}{r}{\small 3.8491} & 
\multicolumn{1}{r}{{\small 1.44}$\times ${\small 10}$^{3}$} & 
\multicolumn{1}{r}{\small 20.8651} & \multicolumn{1}{r}{\small -95.6358} & 
\multicolumn{1}{r}{\small 116.5009} & \multicolumn{1}{r}{{\small 1.01}$%
\times ${\small 10}$^{4}$} \\ 
\multicolumn{1}{l}{} & \multicolumn{1}{l}{} & \multicolumn{1}{l}{} & 
\multicolumn{1}{r}{} & \multicolumn{1}{r}{} & \multicolumn{1}{r}{} & 
\multicolumn{1}{r}{} & \multicolumn{1}{r}{} & \multicolumn{1}{r}{} & 
\multicolumn{1}{r}{} & \multicolumn{1}{r}{} \\ 
\multicolumn{1}{l}{$^{128}${\small Te}} & \multicolumn{1}{l}{\small PHFB} & 
\multicolumn{1}{l}{\small *} & \multicolumn{1}{r}{\small 0.3107} & 
\multicolumn{1}{r}{\small -1.3142} & \multicolumn{1}{r}{\small 1.6249} & 
\multicolumn{1}{r}{\small 22.05} & \multicolumn{1}{r}{\small 9.1133} & 
\multicolumn{1}{r}{\small -42.0190} & \multicolumn{1}{r}{\small 51.1323} & 
\multicolumn{1}{r}{{\small 6.84}$\times ${\small 10}$^{5}$} \\ 
\multicolumn{1}{l}{} & \multicolumn{1}{l}{\small QRPA} & \multicolumn{1}{l}{%
{\small \cite{muto89,hirs96}}} & \multicolumn{1}{r}{\small -1.184} & 
\multicolumn{1}{r}{\small 3.103} & \multicolumn{1}{r}{\small 4.287} & 
\multicolumn{1}{r}{\small 8.36} & \multicolumn{1}{r}{\small -55.0} & 
\multicolumn{1}{r}{\small 248.0} & \multicolumn{1}{r}{\small 303.0} & 
\multicolumn{1}{r}{{\small 4.05}$\times ${\small 10}$^{6}$} \\ 
\multicolumn{1}{l}{} & \multicolumn{1}{l}{\small QRPA} & \multicolumn{1}{l}{%
{\small \cite{tomo91}}} & \multicolumn{1}{r}{\small -0.440} & 
\multicolumn{1}{r}{\small 1.488} & \multicolumn{1}{r}{\small 1.928} & 
\multicolumn{1}{r}{\small 18.58} & \multicolumn{1}{r}{\small -22.200} & 
\multicolumn{1}{r}{\small 100.469} & \multicolumn{1}{r}{\small 122.669} & 
\multicolumn{1}{r}{{\small 1.64}$\times ${\small 10}$^{6}$} \\ 
\multicolumn{1}{l}{} & \multicolumn{1}{l}{\small QRPA} & \multicolumn{1}{l}{%
{\small \cite{pant96}}} & \multicolumn{1}{r}{\small -0.044} & 
\multicolumn{1}{r}{\small 2.437} & \multicolumn{1}{r}{\small 2.480} & 
\multicolumn{1}{r}{\small 14.45} & \multicolumn{1}{r}{} & \multicolumn{1}{r}{
} & \multicolumn{1}{r}{\small 101.233} & \multicolumn{1}{r}{{\small 1.35}$%
\times ${\small 10}$^{6}$} \\ 
\multicolumn{1}{l}{} & \multicolumn{1}{l}{\small FQRPA} & \multicolumn{1}{l}{%
{\small \cite{pant96}}} & \multicolumn{1}{r}{\small 0.391} & 
\multicolumn{1}{r}{\small 1.270} & \multicolumn{1}{r}{\small 0.879} & 
\multicolumn{1}{r}{\small 40.76} & \multicolumn{1}{r}{} & \multicolumn{1}{r}{
} & \multicolumn{1}{r}{\small 43.205} & \multicolumn{1}{r}{{\small 5.78}$%
\times ${\small 10}$^{5}$} \\ 
\multicolumn{1}{l}{} & \multicolumn{1}{l}{\small SM} & \multicolumn{1}{l}{%
{\small \cite{caur07}}} & \multicolumn{1}{r}{\small -0.31} & 
\multicolumn{1}{r}{\small 2.36} & \multicolumn{1}{r}{\small 2.67} & 
\multicolumn{1}{r}{\small 13.42} & \multicolumn{1}{r}{} & \multicolumn{1}{r}{
} & \multicolumn{1}{r}{} & \multicolumn{1}{r}{} \\ 
\multicolumn{1}{l}{} & \multicolumn{1}{l}{\small RQRPA} & \multicolumn{1}{l}{%
{\small \cite{rodi07}}} & \multicolumn{1}{r}{} & \multicolumn{1}{r}{} & 
\multicolumn{1}{r}{{\small 3.23}$_{-0.12}^{+0.12}$} & \multicolumn{1}{r}%
{\small 11.09} & \multicolumn{1}{r}{} & \multicolumn{1}{r}{} & 
\multicolumn{1}{r}{} & \multicolumn{1}{r}{} \\ 
\multicolumn{1}{l}{} & \multicolumn{1}{l}{\small QRPA} & \multicolumn{1}{l}{%
{\small \cite{rodi07}}} & \multicolumn{1}{r}{} & \multicolumn{1}{r}{} & 
\multicolumn{1}{r}{{\small 3.64}$_{-0.13}^{+0.13}$} & \multicolumn{1}{r}%
{\small 9.84} & \multicolumn{1}{r}{} & \multicolumn{1}{r}{} & 
\multicolumn{1}{r}{} & \multicolumn{1}{r}{} \\ 
\multicolumn{1}{l}{} & \multicolumn{1}{l}{\small RQRPA} & \multicolumn{1}{l}{%
{\small \cite{simk08}}} & \multicolumn{1}{r}{} & \multicolumn{1}{r}{} & 
\multicolumn{1}{r}{\small 2.46} & \multicolumn{1}{r}%
{\small 14.56} & \multicolumn{1}{r}{} & \multicolumn{1}{r}{} & 
\multicolumn{1}{r}{} & \multicolumn{1}{r}{} \\ 
\multicolumn{1}{l}{} & \multicolumn{1}{l}{\small RQRPA} & \multicolumn{1}{l}{%
{\small \cite{simk08}}} & \multicolumn{1}{r}{} & \multicolumn{1}{r}{} & 
\multicolumn{1}{r}{\small 3.06} & \multicolumn{1}{r}%
{\small 11.71} & \multicolumn{1}{r}{} & \multicolumn{1}{r}{} & 
\multicolumn{1}{r}{} & \multicolumn{1}{r}{} \\ 
\multicolumn{1}{l}{} & \multicolumn{1}{l}{\small QRPA} & \multicolumn{1}{l}{%
{\small \cite{simk08}}} & \multicolumn{1}{r}{} & \multicolumn{1}{r}{} & 
\multicolumn{1}{r}{\small 3.77} & \multicolumn{1}{r}%
{\small 9.50} & \multicolumn{1}{r}{} & \multicolumn{1}{r}{} & 
\multicolumn{1}{r}{} & \multicolumn{1}{r}{} \\ 
\multicolumn{1}{l}{} & \multicolumn{1}{l}{\small QRPA} & \multicolumn{1}{l}{%
{\small \cite{simk08}}} & \multicolumn{1}{r}{} & \multicolumn{1}{r}{} & 
\multicolumn{1}{r}{\small 4.76} & \multicolumn{1}{r}%
{\small 7.53} & \multicolumn{1}{r}{} & \multicolumn{1}{r}{} & 
\multicolumn{1}{r}{} & \multicolumn{1}{r}{} \\ 
\multicolumn{1}{l}{} & \multicolumn{1}{l}{} & \multicolumn{1}{l}{} & 
\multicolumn{1}{r}{} & \multicolumn{1}{r}{} & \multicolumn{1}{r}{} & 
\multicolumn{1}{r}{} & \multicolumn{1}{r}{} & \multicolumn{1}{r}{} & 
\multicolumn{1}{r}{} & \multicolumn{1}{r}{} \\ 
\multicolumn{1}{l}{$^{130}${\small Te}} & \multicolumn{1}{l}{\small PHFB} & 
\multicolumn{1}{l}{\small *} & \multicolumn{1}{r}{\small 0.4105} & 
\multicolumn{1}{r}{\small -1.8055} & \multicolumn{1}{r}{\small 2.2160} & 
\multicolumn{1}{r}{\small 0.63} & \multicolumn{1}{r}{\small 11.1405} & 
\multicolumn{1}{r}{\small -52.2720} & \multicolumn{1}{r}{\small 63.4125} & 
\multicolumn{1}{r}{{\small 2.18}$\times ${\small 10}$^{7}$} \\ 
\multicolumn{1}{l}{} & \multicolumn{1}{l}{\small QRPA} & \multicolumn{1}{l}{%
{\small \cite{muto89,hirs96}}} & \multicolumn{1}{r}{\small -0.977} & 
\multicolumn{1}{r}{\small 2.493} & \multicolumn{1}{r}{\small 3.470} & 
\multicolumn{1}{r}{\small 0.40} & \multicolumn{1}{r}{\small -48.0} & 
\multicolumn{1}{r}{\small 219.0} & \multicolumn{1}{r}{\small 267.0} & 
\multicolumn{1}{r}{{\small 9.19}$\times ${\small 10}$^{7}$} \\ 
\multicolumn{1}{l}{} & \multicolumn{1}{l}{\small QRPA} & \multicolumn{1}{l}{%
{\small \cite{tomo91}}} & \multicolumn{1}{r}{\small -0.385} & 
\multicolumn{1}{r}{\small 1.289} & \multicolumn{1}{r}{\small 1.674} & 
\multicolumn{1}{r}{\small 0.83} & \multicolumn{1}{r}{\small -19.582} & 
\multicolumn{1}{r}{\small 88.576} & \multicolumn{1}{r}{\small 108.158} & 
\multicolumn{1}{r}{{\small 3.73}$\times ${\small 10}$^{7}$} \\ 
\multicolumn{1}{l}{} & \multicolumn{1}{l}{\small QRPA} & \multicolumn{1}{l}{%
{\small \cite{pant96}}} & \multicolumn{1}{r}{\small -0.009} & 
\multicolumn{1}{r}{\small 2.327} & \multicolumn{1}{r}{\small 2.336} & 
\multicolumn{1}{r}{\small 0.60} & \multicolumn{1}{r}{} & \multicolumn{1}{r}{}
& \multicolumn{1}{r}{\small 92.661} & \multicolumn{1}{r}{{\small 3.19}$%
\times ${\small 10}$^{7}$} \\ 
\multicolumn{1}{l}{} & \multicolumn{1}{l}{\small FQRPA} & \multicolumn{1}{l}{%
{\small \cite{pant96}}} & \multicolumn{1}{r}{\small 0.337} & 
\multicolumn{1}{r}{\small 1.833} & \multicolumn{1}{r}{\small 1.496} & 
\multicolumn{1}{r}{\small 0.93} & \multicolumn{1}{r}{} & \multicolumn{1}{r}{}
& \multicolumn{1}{r}{\small 102.135} & \multicolumn{1}{r}{{\small 3.51}$%
\times ${\small 10}$^{7}$} \\ 
\multicolumn{1}{l}{} & \multicolumn{1}{l}{\small SM} & \multicolumn{1}{l}{%
{\small \cite{caur07}}} & \multicolumn{1}{r}{\small -0.28} & 
\multicolumn{1}{r}{\small 2.13} & \multicolumn{1}{r}{\small 2.41} & 
\multicolumn{1}{r}{\small 0.58} & \multicolumn{1}{r}{} & \multicolumn{1}{r}{}
& \multicolumn{1}{r}{} & \multicolumn{1}{r}{} \\ 
\multicolumn{1}{l}{} & \multicolumn{1}{l}{\small RQRPA} & \multicolumn{1}{l}{%
{\small \cite{rodi07}}} & \multicolumn{1}{r}{} & \multicolumn{1}{r}{} & 
\multicolumn{1}{r}{{\small 2.95}$_{-0.12}^{+0.12}$} & \multicolumn{1}{r}%
{\small 0.47} & \multicolumn{1}{r}{} & \multicolumn{1}{r}{} & 
\multicolumn{1}{r}{} & \multicolumn{1}{r}{} \\ 
\multicolumn{1}{l}{} & \multicolumn{1}{l}{\small QRPA} & \multicolumn{1}{l}{%
{\small \cite{rodi07}}} & \multicolumn{1}{r}{} & \multicolumn{1}{r}{} & 
\multicolumn{1}{r}{{\small 3.26}$_{-0.12}^{+0.12}$} & \multicolumn{1}{r}%
{\small 0.43} & \multicolumn{1}{r}{} & \multicolumn{1}{r}{} & 
\multicolumn{1}{r}{} & \multicolumn{1}{r}{} \\ 
\multicolumn{1}{l}{} & \multicolumn{1}{l}{\small RQRPA} & \multicolumn{1}{l}{%
{\small \cite{simk08}}} & \multicolumn{1}{r}{} & \multicolumn{1}{r}{} & 
\multicolumn{1}{r}{\small 2.27} & \multicolumn{1}{r}%
{\small 0.61} & \multicolumn{1}{r}{} & \multicolumn{1}{r}{} & 
\multicolumn{1}{r}{} & \multicolumn{1}{r}{} \\ 
\multicolumn{1}{l}{} & \multicolumn{1}{l}{\small RQRPA} & \multicolumn{1}{l}{%
{\small \cite{simk08}}} & \multicolumn{1}{r}{} & \multicolumn{1}{r}{} & 
\multicolumn{1}{r}{\small 2.84} & \multicolumn{1}{r}%
{\small 0.49} & \multicolumn{1}{r}{} & \multicolumn{1}{r}{} & 
\multicolumn{1}{r}{} & \multicolumn{1}{r}{} \\ 
\multicolumn{1}{l}{} & \multicolumn{1}{l}{\small QRPA} & \multicolumn{1}{l}{%
{\small \cite{simk08}}} & \multicolumn{1}{r}{} & \multicolumn{1}{r}{} & 
\multicolumn{1}{r}{\small 3.38} & \multicolumn{1}{r}%
{\small 0.41} & \multicolumn{1}{r}{} & \multicolumn{1}{r}{} & 
\multicolumn{1}{r}{} & \multicolumn{1}{r}{} \\ 
\multicolumn{1}{l}{} & \multicolumn{1}{l}{\small QRPA} & \multicolumn{1}{l}{%
{\small \cite{simk08}}} & \multicolumn{1}{r}{} & \multicolumn{1}{r}{} & 
\multicolumn{1}{r}{\small 4.26} & \multicolumn{1}{r}%
{\small 0.33} & \multicolumn{1}{r}{} & \multicolumn{1}{r}{} & 
\multicolumn{1}{r}{} & \multicolumn{1}{r}{} \\ 
\multicolumn{1}{l}{} & \multicolumn{1}{l}{} & \multicolumn{1}{l}{} & 
\multicolumn{1}{r}{} & \multicolumn{1}{r}{} & \multicolumn{1}{r}{} & 
\multicolumn{1}{r}{} & \multicolumn{1}{r}{} & \multicolumn{1}{r}{} & 
\multicolumn{1}{r}{} & \multicolumn{1}{r}{} \\ 
\multicolumn{1}{l}{$^{150}${\small Nd}} & \multicolumn{1}{l}{\small PHFB} & 
\multicolumn{1}{l}{\small *} & \multicolumn{1}{r}{\small 0.3039} & 
\multicolumn{1}{r}{\small -1.3119} & \multicolumn{1}{r}{\small 1.6158} & 
\multicolumn{1}{r}{\small 19.85} & \multicolumn{1}{r}{\small 8.0695} & 
\multicolumn{1}{r}{\small -37.3676} & \multicolumn{1}{r}{\small 45.4371} & 
\multicolumn{1}{r}{{\small 6.79}$\times ${\small 10}$^{5}$} \\ 
\multicolumn{1}{l}{} & \multicolumn{1}{l}{\small QRPA} & \multicolumn{1}{l}{%
{\small \cite{muto89,hirs96}}} & \multicolumn{1}{r}{\small -1.821} & 
\multicolumn{1}{r}{\small 4.254} & \multicolumn{1}{r}{\small 6.075} & 
\multicolumn{1}{r}{\small 5.28} & \multicolumn{1}{r}{\small -78.0} & 
\multicolumn{1}{r}{\small 344.0} & \multicolumn{1}{r}{\small 422.0} & 
\multicolumn{1}{r}{{\small 6.31}$\times ${\small 10}$^{6}$} \\ 
\multicolumn{1}{l}{} & \multicolumn{1}{l}{\small QRPA} & \multicolumn{1}{l}{%
{\small \cite{tomo91}}} & \multicolumn{1}{r}{\small -0.629} & 
\multicolumn{1}{r}{\small 1.989} & \multicolumn{1}{r}{\small 2.618} & 
\multicolumn{1}{r}{\small 12.25} & \multicolumn{1}{r}{\small -28.385} & 
\multicolumn{1}{r}{\small 124.700} & \multicolumn{1}{r}{\small 153.085} & 
\multicolumn{1}{r}{{\small 2.29}$\times ${\small 10}$^{6}$} \\ 
\multicolumn{1}{l}{} & \multicolumn{1}{l}{\small pSU(3)} & 
\multicolumn{1}{l}{{\small \cite{jghi95}}} & \multicolumn{1}{r}{} & 
\multicolumn{1}{r}{} & \multicolumn{1}{r}{\small 1.570} & \multicolumn{1}{r}%
{\small 20.43} & \multicolumn{1}{r}{} & \multicolumn{1}{r}{} & 
\multicolumn{1}{r}{} & \multicolumn{1}{r}{} \\ 
\multicolumn{1}{l}{} & \multicolumn{1}{l}{\small RQRPA} & \multicolumn{1}{l}{%
{\small \cite{rodi07}}} & \multicolumn{1}{r}{} & \multicolumn{1}{r}{} & 
\multicolumn{1}{r}{{\small 4.16}$_{-0.16}^{+0.16}$} & \multicolumn{1}{r}%
{\small 7.71} & \multicolumn{1}{r}{} & \multicolumn{1}{r}{} & 
\multicolumn{1}{r}{} & \multicolumn{1}{r}{} \\ 
\multicolumn{1}{l}{} & \multicolumn{1}{l}{\small QRPA} & \multicolumn{1}{l}{%
{\small \cite{rodi07}}} & \multicolumn{1}{r}{} & \multicolumn{1}{r}{} & 
\multicolumn{1}{r}{{\small 4.74}$_{-0.20}^{+0.20}$} & \multicolumn{1}{r}%
{\small 6.77} & \multicolumn{1}{r}{} & \multicolumn{1}{r}{} & 
\multicolumn{1}{r}{} & \multicolumn{1}{r}{} \\ 
\multicolumn{1}{l}{} & \multicolumn{1}{l}{} & \multicolumn{1}{l}{} & 
\multicolumn{1}{r}{} & \multicolumn{1}{r}{} & \multicolumn{1}{r}{} & 
\multicolumn{1}{r}{} & \multicolumn{1}{r}{} & \multicolumn{1}{r}{} & 
\multicolumn{1}{r}{} & \multicolumn{1}{r}{} \\ \hline\hline
\end{tabular}
\end{table}

\begin{table}
\caption{Effect of variation in $\zeta _{qq}$
on NTMEs $M_{\alpha }$ $(\alpha =F,GT,Fh$ and $GTh)$ of $\left( \beta
^{-}\beta ^{-}\right) _{0\nu }$ decay.}

\begin{tabular}{llllllllllllllll}
\hline\hline
& $\zeta _{qq}$ & {\small 0.00} & {\small 0.20} & {\small 0.40} & {\small %
0.60} & {\small 0.80} & {\small 0.90} & {\small 0.95} & {\small 1.00} & 
{\small 1.05} & {\small 1.10} & {\small 1.20} & {\small 1.30} & {\small 1.40}
& {\small 1.50} \\ \hline
&  &  &  &  &  &  &  &  &  &  &  &  &  &  &  \\ 
$^{94}${\small Zr} & $\left| M_{F}\right| $ & {\small 0.9525} & {\small %
0.9138} & {\small 0.9332} & {\small 0.9325} & {\small 0.8380} & {\small %
0.3889} & {\small 0.3450} & {\small 0.3433} & {\small 0.3118} & {\small %
0.2755} & {\small 0.0716} & {\small 0.0777} & {\small 0.0419} & {\small %
0.1962} \\ 
& $\left| M_{GT}\right| $ & {\small 4.1580} & {\small 4.0673} & {\small %
4.1432} & {\small 4.1588} & {\small 3.8157} & {\small 1.8974} & {\small %
1.6863} & {\small 1.6912} & {\small 1.5779} & {\small 1.4556} & {\small %
0.3196} & {\small 0.3535} & {\small 0.2095} & {\small 1.0231} \\ 
& $\left| M_{Fh}\right| $ & {\small 26.4106} & {\small 25.7395} & {\small %
26.1951} & {\small 26.2492} & {\small 24.1958} & {\small 12.3099} & {\small %
11.0816} & {\small 11.1290} & {\small 10.3310} & {\small 9.3615} & {\small %
2.5563} & {\small 2.8608} & {\small 1.9605} & {\small 6.1134} \\ 
& $\left| M_{GTh}\right| $ & {\small 120.702} & {\small 118.031} & {\small %
120.066} & {\small 120.406} & {\small 111.251} & {\small 56.9742} & {\small %
51.1542} & {\small 51.5522} & {\small 47.9816} & {\small 43.6137} & {\small %
11.4948} & {\small 12.8413} & {\small 8.8149} & {\small 27.9369} \\ 
&  &  &  &  &  &  &  &  &  &  &  &  &  &  &  \\ 
$^{96}${\small Zr} & $\left| M_{F}\right| $ & {\small 1.2254} & {\small %
1.2627} & {\small 1.2513} & {\small 1.2397} & {\small 0.9641} & {\small %
0.6311} & {\small 0.4358} & {\small 0.2576} & {\small 0.1347} & {\small %
0.0773} & {\small 0.0302} & {\small 0.0772} & {\small 0.0730} & {\small %
0.0624} \\ 
& $\left| M_{GT}\right| $ & {\small 5.3171} & {\small 5.4406} & {\small %
5.4482} & {\small 5.4491} & {\small 4.4492} & {\small 3.0011} & {\small %
2.0655} & {\small 1.1958} & {\small 0.6002} & {\small 0.3336} & {\small %
0.1201} & {\small 0.3148} & {\small 0.3050} & {\small 0.2669} \\ 
& $\left| M_{Fh}\right| $ & {\small 33.4814} & {\small 34.2490} & {\small %
34.2199} & {\small 34.1506} & {\small 28.0203} & {\small 19.4408} & {\small %
13.8870} & {\small 8.5271} & {\small 4.5763} & {\small 2.6647} & {\small %
1.0640} & {\small 2.5368} & {\small 2.5209} & {\small 2.2641} \\ 
& $\left| M_{GTh}\right| $ & {\small 152.749} & {\small 156.080} & {\small %
156.213} & {\small 156.145} & {\small 128.921} & {\small 89.3919} & {\small %
63.4324} & {\small 38.4867} & {\small 20.3309} & {\small 11.7475} & {\small %
4.6909} & {\small 11.1870} & {\small 11.1178} & {\small 9.9883} \\ 
&  &  &  &  &  &  &  &  &  &  &  &  &  &  &  \\ 
$^{98}${\small Mo} & $\left| M_{F}\right| $ & {\small 1.2471} & {\small %
1.2890} & {\small 1.3440} & {\small 1.2762} & {\small 0.5080} & {\small %
0.2813} & {\small 0.6198} & {\small 0.6159} & {\small 0.1425} & {\small %
0.0799} & {\small 0.0420} & {\small 0.0478} & {\small 0.1829} & {\small %
0.1325} \\ 
& $\left| M_{GT}\right| $ & {\small 5.2875} & {\small 5.4677} & {\small %
5.6704} & {\small 5.4994} & {\small 2.3544} & {\small 1.3173} & {\small %
2.8606} & {\small 2.7502} & {\small 0.5978} & {\small 0.3480} & {\small %
0.1970} & {\small 0.2278} & {\small 0.8807} & {\small 0.6494} \\ 
& $\left| M_{Fh}\right| $ & {\small 33.4621} & {\small 34.5055} & {\small %
35.7334} & {\small 34.6749} & {\small 15.6358} & {\small 9.1876} & {\small %
18.2205} & {\small 17.2313} & {\small 4.3304} & {\small 2.4780} & {\small %
1.2824} & {\small 1.4689} & {\small 5.2861} & {\small 3.9076} \\ 
& $\left| M_{GTh}\right| $ & {\small 151.618} & {\small 156.392} & {\small %
161.819} & {\small 157.628} & {\small 71.7167} & {\small 42.0018} & {\small %
83.7605} & {\small 78.8982} & {\small 19.4513} & {\small 11.3293} & {\small %
6.1742} & {\small 7.1270} & {\small 25.2409} & {\small 19.1005} \\ 
&  &  &  &  &  &  &  &  &  &  &  &  &  &  &  \\ 
$^{100}${\small Mo} & $\left| M_{F}\right| $ & {\small 1.2780} & {\small %
1.4364} & {\small 1.4561} & {\small 1.5051} & {\small 1.5178} & {\small %
0.6441} & {\small 0.8557} & {\small 0.6013} & {\small 0.3241} & {\small %
0.2992} & {\small 0.3380} & {\small 0.4183} & {\small 0.1560} & {\small %
0.1741} \\ 
& $\left| M_{GT}\right| $ & {\small 5.7255} & {\small 6.2099} & {\small %
6.3320} & {\small 6.5392} & {\small 6.6479} & {\small 3.0258} & {\small %
3.9287} & {\small 2.6505} & {\small 1.3574} & {\small 1.2403} & {\small %
1.3153} & {\small 1.6460} & {\small 0.7130} & {\small 0.7717} \\ 
& $\left| M_{Fh}\right| $ & {\small 35.3409} & {\small 38.4610} & {\small %
39.1496} & {\small 40.3860} & {\small 40.9767} & {\small 19.6010} & {\small %
25.4105} & {\small 17.9096} & {\small 9.5889} & {\small 8.9123} & {\small %
9.5424} & {\small 11.3697} & {\small 4.4923} & {\small 4.6833} \\ 
& $\left| M_{GTh}\right| $ & {\small 161.769} & {\small 174.981} & {\small %
178.291} & {\small 183.884} & {\small 186.830} & {\small 90.2545} & {\small %
116.663} & {\small 81.6304} & {\small 43.7223} & {\small 40.4320} & {\small %
42.7095} & {\small 50.5883} & {\small 21.3888} & {\small 21.5911} \\ 
&  &  &  &  &  &  &  &  &  &  &  &  &  &  &  \\ 
$^{104}${\small Ru} & $\left| M_{F}\right| $ & {\small 1.6121} & {\small %
1.5740} & {\small 1.5482} & {\small 1.5771} & {\small 0.6685} & {\small %
0.6570} & {\small 0.5946} & {\small 0.4684} & {\small 0.3277} & {\small %
0.2351} & {\small 0.1826} & {\small 0.0241} & {\small 0.1693} & {\small %
0.1667} \\ 
& $\left| M_{GT}\right| $ & {\small 7.3361} & {\small 7.2600} & {\small %
7.2490} & {\small 7.3978} & {\small 3.0294} & {\small 2.8126} & {\small %
2.4618} & {\small 1.8787} & {\small 1.2527} & {\small 0.8447} & {\small %
0.6502} & {\small 0.0810} & {\small 0.3864} & {\small 0.3802} \\ 
& $\left| M_{Fh}\right| $ & {\small 43.9292} & {\small 43.4102} & {\small %
43.2413} & {\small 44.1136} & {\small 19.5841} & {\small 19.3482} & {\small %
17.4213} & {\small 13.8055} & {\small 9.6922} & {\small 6.7778} & {\small %
5.2601} & {\small 0.6816} & {\small 3.8479} & {\small 3.7724} \\ 
& $\left| M_{GTh}\right| $ & {\small 201.252} & {\small 199.327} & {\small %
199.057} & {\small 203.139} & {\small 90.9183} & {\small 88.5215} & {\small %
79.2869} & {\small 62.7040} & {\small 43.9749} & {\small 30.8020} & {\small %
23.8972} & {\small 2.9638} & {\small 14.0488} & {\small 13.7353} \\ 
&  &  &  &  &  &  &  &  &  &  &  &  &  &  &  \\ 
$^{110}${\small Pd} & $\left| M_{F}\right| $ & {\small 1.7381} & {\small %
1.7238} & {\small 1.7485} & {\small 1.7469} & {\small 1.3849} & {\small %
0.9328} & {\small 0.7303} & {\small 0.6880} & {\small 0.5537} & {\small %
0.3456} & {\small 0.2808} & {\small 0.1851} & {\small 0.1458} & {\small %
0.1118} \\ 
& $\left| M_{GT}\right| $ & {\small 8.4105} & {\small 8.3851} & {\small %
8.4792} & {\small 8.4913} & {\small 6.5005} & {\small 4.2640} & {\small %
3.3472} & {\small 3.1611} & {\small 2.6277} & {\small 1.7438} & {\small %
1.3719} & {\small 0.8580} & {\small 0.6526} & {\small 0.5252} \\ 
& $\left| M_{Fh}\right| $ & {\small 49.0177} & {\small 48.8272} & {\small %
49.4187} & {\small 49.4697} & {\small 40.1148} & {\small 28.3224} & {\small %
22.6514} & {\small 20.8651} & {\small 16.2553} & {\small 10.5955} & {\small %
8.4977} & {\small 5.4096} & {\small 4.3595} & {\small 3.6178} \\ 
& $\left| M_{GTh}\right| $ & {\small 225.933} & {\small 225.220} & {\small %
227.828} & {\small 228.150} & {\small 185.198} & {\small 129.798} & {\small %
103.531} & {\small 95.6358} & {\small 75.4985} & {\small 49.9589} & {\small %
40.3671} & {\small 26.0640} & {\small 21.4753} & {\small 18.2737} \\ 
&  &  &  &  &  &  &  &  &  &  &  &  &  &  &  \\ 
$^{128}${\small Te} & $\left| M_{F}\right| $ & {\small 1.3400} & {\small %
1.3207} & {\small 1.3410} & {\small 0.6911} & {\small 0.3791} & {\small %
0.3318} & {\small 0.3096} & {\small 0.3107} & {\small 0.2643} & {\small %
0.2165} & {\small 0.0207} & {\small 0.0033} & {\small 0.0002}$^{\dagger }$ & 
{\small 0.0004} \\ 
& $\left| M_{GT}\right| $ & {\small 5.8530} & {\small 5.7535} & {\small %
5.8369} & {\small 3.1351} & {\small 1.6661} & {\small 1.4220} & {\small %
1.3133} & {\small 1.3142} & {\small 1.1136} & {\small 0.9261} & {\small %
0.0962} & {\small 0.0147} & {\small 0.0008}$^{\dagger }$ & {\small 0.0015}
\\ 
& $\left| M_{Fh}\right| $ & {\small 35.4843} & {\small 35.3388} & {\small %
35.8018} & {\small 18.7882} & {\small 10.7613} & {\small 9.6002} & {\small %
9.0907} & {\small 9.1133} & {\small 8.0909} & {\small 6.8243} & {\small %
0.7938} & {\small 0.1577} & {\small 0.0138}$^{\dagger }$ & {\small 0.0249}
\\ 
& $\left| M_{GTh}\right| $ & {\small 166.140} & {\small 165.460} & {\small %
167.632} & {\small 88.4856} & {\small 50.3124} & {\small 44.4687} & {\small %
41.9310} & {\small 42.0190} & {\small 37.2069} & {\small 31.6283} & {\small %
3.9705} & {\small 0.7864} & {\small 0.0680}$^{\dagger }$ & {\small 0.1220}
\\ 
&  &  &  &  &  &  &  &  &  &  &  &  &  &  &  \\ 
$^{130}${\small Te} & $\left| M_{F}\right| $ & {\small 1.2162} & {\small %
1.2339} & {\small 1.2371} & {\small 1.1357} & {\small 0.5663} & {\small %
0.5283} & {\small 0.4628} & {\small 0.4105} & {\small 0.3690} & {\small %
0.3610} & {\small 0.3206} & {\small 0.2697} & {\small 0.2570} & {\small %
0.1062} \\ 
& $\left| M_{GT}\right| $ & {\small 5.3383} & {\small 5.4108} & {\small %
5.4179} & {\small 5.0086} & {\small 2.5392} & {\small 2.3577} & {\small %
2.0562} & {\small 1.8055} & {\small 1.6091} & {\small 1.5650} & {\small %
1.3646} & {\small 1.1307} & {\small 1.0735} & {\small 0.4562} \\ 
& $\left| M_{Fh}\right| $ & {\small 32.4773} & {\small 32.9039} & {\small %
33.0896} & {\small 30.4365} & {\small 14.9999} & {\small 14.1131} & {\small %
12.5245} & {\small 11.1405} & {\small 10.1820} & {\small 9.9612} & {\small %
8.8699} & {\small 7.7887} & {\small 7.4791} & {\small 3.3935} \\ 
& $\left| M_{GTh}\right| $ & {\small 151.956} & {\small 153.963} & {\small %
154.832} & {\small 142.316} & {\small 70.4659} & {\small 66.3531} & {\small %
58.8479} & {\small 52.2720} & {\small 47.6396} & {\small 46.5332} & {\small %
41.2186} & {\small 35.9060} & {\small 34.4146} & {\small 15.8394} \\ 
&  &  &  &  &  &  &  &  &  &  &  &  &  &  &  \\ 
$^{150}${\small Nd} & $\left| M_{F}\right| $ & {\small 1.7574} & {\small %
1.7866} & {\small 1.7941} & {\small 1.7618} & {\small 1.1265} & {\small %
0.8929} & {\small 0.4816} & {\small 0.3039} & {\small 0.2466} & {\small %
0.2087} & {\small 0.1809} & {\small 0.1258} & {\small 0.1226} & {\small %
0.0981} \\ 
& $\left| M_{GT}\right| $ & {\small 8.1997} & {\small 8.3249} & {\small %
8.3573} & {\small 8.1859} & {\small 5.2625} & {\small 4.1539} & {\small %
2.1505} & {\small 1.3119} & {\small 1.0927} & {\small 0.9657} & {\small %
0.9218} & {\small 0.7280} & {\small 0.6640} & {\small 0.5203} \\ 
& $\left| M_{Fh}\right| $ & {\small 49.9349} & {\small 50.7036} & {\small %
50.9731} & {\small 50.0155} & {\small 31.8413} & {\small 24.7475} & {\small %
13.0967} & {\small 8.0695} & {\small 6.5722} & {\small 5.6055} & {\small %
5.2347} & {\small 4.0764} & {\small 3.9110} & {\small 3.2948} \\ 
& $\left| M_{GTh}\right| $ & {\small 234.218} & {\small 237.784} & {\small %
238.980} & {\small 234.268} & {\small 149.171} & {\small 116.206} & {\small %
61.0345} & {\small 37.3676} & {\small 30.6928} & {\small 26.4558} & {\small %
25.3258} & {\small 20.5210} & {\small 19.8597} & {\small 17.1640} \\ 
&  &  &  &  &  &  &  &  &  &  &  &  &  &  &  \\ \hline\hline
\end{tabular}
\footnotetext{$^{\dagger }$denotes $\zeta _{qq}=1.41.$}
\end{table}

\begin{table}
\caption{Ratios $D_{\alpha }$ for $^{94,96}$Zr, $%
^{98,100}$Mo, $^{104}$Ru, $^{110}$Pd, $^{128,130}$Te and $^{150}$Nd isotopes.}

\begin{tabular}{llllllllll}
\hline\hline
Ratios & $^{94}$Zr & $^{96}$Zr & $^{98}$Mo & $^{100}$Mo & $^{104}$Ru & $%
^{110}$Pd & $^{128}$Te & $^{130}$Te & $^{150}$Nd \\ \hline
&  &  &  &  &  &  &  &  &  \\ 
$D_{F}$ & 2.77 & 4.76 & 2.02 & 2.13 & 3.44 & 2.53 & 4.31 & 2.96 & 5.78 \\ 
&  &  &  &  &  &  &  &  &  \\ 
$D_{GT}$ & 2.46 & 4.45 & 1.92 & 2.16 & 3.90 & 2.66 & 4.45 & 2.96 & 6.25 \\ 
&  &  &  &  &  &  &  &  &  \\ 
$D_{Fh}$ & 2.37 & 3.93 & 1.94 & 1.97 & 3.18 & 2.35 & 3.89 & 2.92 & 6.19 \\ 
&  &  &  &  &  &  &  &  &  \\ 
$D_{GTh}$ & 2.34 & 3.97 & 1.92 & 1.98 & 3.21 & 2.36 & 3.95 & 2.91 & 6.27 \\ 
&  &  &  &  &  &  &  &  &  \\ 
$D_{2\nu }$ & 2.29 & 3.70 & 1.86 & 2.33 & 5.47 & 3.14 & 4.26 & 2.89 & 5.94
\\ 
&  &  &  &  &  &  &  &  &  \\ \hline\hline
\end{tabular}
\end{table}

\end{document}